\documentclass{aastex}
\usepackage{spr-astr-addons}
\usepackage{url}
\usepackage{morefloats}

\RequirePackage{color}

\providecommand{\abs}[1]{\lvert#1\rvert}

\begin{document}

\title{A systematic cross-search for radio/infrared counterparts of XMM-{\it Newton} sources}
\slugcomment{}
%% Running heads
\shorttitle{Radio/infrared counterparts of XMM-{\it Newton} sources}
\shortauthors{Combi et al.}

\author{J.~A. Combi\altaffilmark{1,5,6}} 
\affil{$^1$ Facultad de Ciencias Astron\'omicas y Geof\'{\i}sicas, 
Universidad Nacional de La Plata, Paseo del Bosque, B1900FWA La Plata, Argentina}

\and 

\author{J.~F. Albacete Colombo\altaffilmark{1,2}} \and
\affil{$^2$ Centro Universitario Regional Zona Atl\'antica (CURZA). 
Universidad Nacional del COMAHUE, Monse\~nor Esandi y Ayacucho (8500),
Viedma (Rio Negro), Argentina}

\and

\author{L. Pellizza\altaffilmark{3}} \and
\affil{$^3$ Instituto de Astronom\'{\i}a y F\'{\i}sica del Espacio, C.C. 67, Suc. 28, 
1428, Buenos Aires, Argentina}

\and

\author{J. L\'opez-Santiago\altaffilmark{4}} 
\affil{$^4$ Departamento de Astrof\'{\i}sica y Ciencias de la Atm\'osfera, Universidad 
Complutense de Madrid, E-28040, Madrid, Spain}

\and

\author{G.~E. Romero\altaffilmark{1,5}} 
\affil{$^5$ Instituto Argentino de Radioastronom\'{\i}a (CCT La Plata, CONICET), C.C.5, 
(1894) Villa Elisa, Buenos Aires, Argentina.}

\and

\author{J.  Mart\'{\i}\altaffilmark{6} and A.~J. Mu\~noz-Arjonilla\altaffilmark{6} and
E. S\'anchez-Ayaso\altaffilmark{6}}
\affil{$^6$ Departamento de F\'{\i}sica (EPS), Universidad de Ja\'en,
Campus Las Lagunillas s/n, A3, 23071 Ja\'en, Spain}

\and

\author{P.~L. Luque-Escamilla\altaffilmark{6,7}} 
\affil{$^7$ Departamento de Ingenier\'{\i}a Mec\'anica y Minera, 
Escuela Polit\'ecnica Superior, Universidad de Ja\'en, 
Campus Las Lagunillas s/n, A3, 23071 Ja\'en, Spain.}

%\and
%
%\author{A.~J. Mu\~noz-Arjonilla\altaffilmark{6}} 
%\affil{Departamento de F\'{\i}sica (EPS), Universidad de Ja\'en,
%Campus Las Lagunillas s/n, A3, 23071 Ja\'en, Spain}

\and

\author{J.~R. S\'anchez-Sutil\altaffilmark{8}} 
\affil{$^8$ Grupo de Investigaci\'on FQM-322, Universidad de Ja\'en, 
Campus Las Lagunillas s/n, A3, 23071 Ja\'en, Spain.}

%\and
%
%\author{E. S\'anchez-Ayaso\altaffilmark{6}}
%\affil{Departamento de F\'{\i}sica (EPS), Universidad de Ja\'en,
%Campus Las Lagunillas s/n, A3, 23071 Ja\'en, Spain}
%\email{\emaila}

%\altaffiltext{1}{Facultad de Ciencias Astron\'omicas y Geof\'{\i}sicas, 
%Universidad Nacional de La Plata, Paseo del Bosque, B1900FWA La Plata, Argentina}
%\altaffiltext{2}{Second Alternate Affilation.}
%\altaffiltext{3}{Third Alternate Affilation.}

\begin{abstract}
%We performed a cross-correlation with very restrictive selection criteria for radio, 
We present a catalog of cross-correlated radio, infrared and X-ray sources using a
very restrictive selection criteria with an IDL-based code developed by us. The significance 
of the observed coincidences was evaluated through Monte Carlo simulations of synthetic 
sources following a well-tested protocol. We found 3320 coincident radio/X-ray 
sources with a high statistical significance characterized by the sum of error-weighted 
coordinate differences. For 997 of them, 2MASS counterparts were found. The 
percentage of chance coincidences is less than 1\%. 
X-ray hardness ratios of well-known populations of objects were 
used to provide a crude representation of their X-ray spectrum and to make a 
preliminary diagnosis of the possible nature of unidentified X-ray sources. The results 
support the fact that the X-ray sky is largely dominated by Active Galactic Nuclei at high
galactic latitudes ($|b| \geq 10 \degr$). At low galactic latitudes ($|b| \leq 10 \degr$) 
most of unidentified X-ray sources ($\sim$ 94$\%$) lie at $|b| \leq 2 \degr$. This 
result suggests that most of the unidentified sources found toward the Milky Way plane 
are galactic objects. %This study intends to provide a useful preliminary diagnosis 
%tool that allows to deepen into our understanding of high-energy X-ray sources 
%with radio and infrared counterpart.
Well-known and unidentified sources were classified in different tables with their 
corresponding radio/infrared and X-ray properties. These tables are intended as a 
useful tool for researchers interested in particular identifications.

\end{abstract}


%\section*{}
%\label{sec:intro}

\section{Introduction}

Understanding the nature of the galactic and extragalactic X-ray sources and the
physical phenomena that produce the emission is one of the main goals of
modern astrophysics. Astrophysical systems emitting at high energies are
usually complex, comprising objects or media capable of emitting over different
ranges of the electromagnetic spectrum \citep[e.g.][]{fabbiano06}. 
%(e.g., Fabbiano 2006 and references therein). 
Multiwavelength analyses are needed to disentangle and
characterize the different components of such astrophysical systems.

The last generation of high-sensitivity X-ray observatories, like {\em Chandra}
and {\em XMM-Newton}, have led to the detection of large samples of X-ray
sources \citep{watson09}. %(e.g., Watson et~al. 2009). 
These observatories provide a new and more
detailed view of the X-ray Universe having an impact on our understanding of
different populations of high-energy objects. In past years several studies of
X-ray sources, at high and middle galactic latitude, have been carried out to
investigate various statistical properties of X-ray sources \citep{sever03,mateos05,
carrera07,caccianiga07,mateos08,della04,della08,ebrero09}.
%(Severgnini et~al. 2003; Mateos et~al. 2005; Carrera et~al. 2007; Caccianiga et~al. 
%2007; Mateos et~al. 2008; Della Ceca et~al. 2004, 2008; Ebrero et~al. 2009). 
Type 1 AGNs (Seyferts) seem to be the dominating population, but type 2 AGNs,
absorption-line galaxies, clusters of galaxies, and stars also contribute to
the total number of detected sources \citep{barcons02,barcons07,barcons03,
mice07,lopez07,feruglio08,novara09}.
%(Barcons et~al. 2002, 2007; Barcons \&
%Negueruela 2003; Micela et~al. 2007; L\'opez-Santiago et~al. 2007; Feruglio
%et~al. 2008; Novara et~al. 2009). 
The size of these dataset (over
$10^5$ sources) prevents detailed multiwavelength studies of any statistically
meaningful subsample of them. The only way to expand the knowledge of their
properties to other regions of the electromagnetic spectrum is to
cross-correlate these samples with surveys at other wavelengths.
Cross-correlation is a powerful method in data analysis to compare the
similarity and to quantitatively measure the relationship between two data
groups. A large volume of papers describing several methods exists 
\citep[e.g.][]{peebles74,seldner77,furenlid90,boyle95}, 
%(e.g. Peebles 1974; Seldner \& Peebles 1977; Furenlid \& Furenlid 1990; Boyle et~al.
%1995; Brown et~al. 2009, Shone et~al. 2009), 
and many important applications have been developed since the first work 
by \citep{woodward95}. %Woodward (1955)

The cross-correlation of X-ray surveys with others at different wavelengths,
and the analysis of their properties along the whole electromagnetic spectrum
is important in several ways. First, it would provide clues on the nature of
individual sources and their emission. Second, it would allow to acquire a
deeper and more accurate insight into the statistical properties of
different X-ray source populations, particularly for rare objects. It would also
provide the basis for statistical modeling of such populations, which might
give insight into the origin and evolution of systems emitting at high energies
(e.g., Fabbiano 2006). Finally, it might reveal the existence of new classes of
objects. Unidentified sources are particularly interesting for this last point.

The XMM-Newton Serendipitous Source Catalogue \citep[2XMMi;][]{watson09}
%(2XMMi, Watson et~al. 2009) 
is the largest survey of X-ray sources available at present, providing nearly
homogeneous data for almost $2.5 \times 10^5$ sources at arcsecond angular
resolution. Most of these sources
have been identified with others at longer wavelengths, while many others
remain unidentified. \citet{watson09} %Watson et~al. (2009) 
provide cross-identification of the
2XMMi sources with many other catalogs, including NVSS, 2MASS, USNO, etc.
However, they do not investigate the statistical properties of the correlated
samples. A recent study by \citet{flesch10} %\citep[see also][]{flesh04} 
provides a thorough investigation
on the correlation of 2XMMi sources (and also of those from other XMM and
Chandra catalogues) with different classes of optical objects. They also
correlate the optical sources with radio catalogs, although they avoid a direct
cross-identification of X-ray with radio sources.

In this paper, we present a cross-identification of the 2XMMi sources with
different radio catalogues covering the whole sky, the NRAO-VLA Sky Survey
\citep[NVSS;][]{condon98}, the Sydney University Molonglo Sky Survey
\citep[SUMSS;][]{mauch07} and the Molonglo Galactic Plane Survey 2
\citep[MGPS-2;][]{green99}.
%(NVSS, Condon 1998), (SUMSS; Mauch et~ al. 2007) (MGPS-2; Green et~al. 1999). 
These catalogues were chosen because of their complementary sky coverage
and their homogeneity (both internal and between catalogues). These properties
allow us to perform a statistically significant analysis of X-ray source
populations and their properties at radio wavelengths. We also perform a
cross-correlation with the 2MASS Catalogue \citep{cutri03}. %(Cutri et~al. 2003). 
Using restrictive selection criteria, we assess the reliability of these
cross-correlations and we use characteristics of well-known X-ray populations
to make preliminary diagnosis of the nature of unidentified X-ray sources with
radio and infrared counterparts. The structure of the paper is as follows: in
Sect.~\ref{correlation} we describe the cross-correlation analysis and strategy
to compute the positional correlation of sources. The main results and
discussion are presented in Sect.~\ref{results}. Finally, we summarize the main
conclusions in Sect.~\ref{summary}. 

\section{Cross-correlation analysis} 
\label{correlation}

\subsection{The catalogues} 

To perform a positional cross-identification between radio and X-ray sources 
we have used the NRAO VLA Sky Survey (NVSS\footnote{http://www.cv.nrao.edu/nvss/}) 
catalog \citep{condon98}, the %(Condon et al. 1998), the
Sydney University Molonglo Sky Survey
\citep[SUMSS\footnote{http://www.astrop.physics.usyd.edu.au/sumsscat/};][]{mauch07}, 
%(SUMSS\footnote{http://www.astrop.physics.usyd.edu.au/sumsscat/}) (Mauch et al. 2007),
 and the Molonglo Galactic Plane Survey
\citep[MGPS2\footnote{http://www.astrop.physics.usyd.edu.au/MGPS/ORIGINALS/};][]{green99}
%(MGPS2\footnote{http://www.astrop.physics.usyd.edu.au/MGPS/ORIGINALS/}) (Green et al. 1999)
 with the XMM-{\it Newton} Serendipitous Source Catalog (Second
Version: 2XMM\footnote{http://heasarc.gsfc.nasa.gov/FTP/xmm/data/catalogs/\\ 2XMMcat$_{-}$v1.0.fits.gz} 
, 2007). The three radio surveys cover the whole sky.

The NVSS catalog (at 1420 MHz) contains 1773484 sources. It covers the entire
sky north of $-40 \degr$ declination. The rms positional uncertainties in
RA and DEC vary from $1\arcsec$ for relatively strong ($S$ $\geq$ 15 mJy) point
sources, to $7 \arcsec$ for the faintest ($S$$\leq$2.3 mJy) detectable sources.
The SUMSS catalog (at 843 MHz) consists of 210412 radio sources. It covers
the southern sky with $\delta \leq -30 \degr$. Positions in the catalog are
accurate within $1 \arcsec$ to $2 \arcsec$ for sources with peak brightness
$\geq$ 20 mJy/beam and are always better than $10 \arcsec$. The MGPS2 is a
radio continuum survey carried out with the Molonglo Observatory Synthesis
Telescope (MOST) at 843 MHz with a resolution of $45 \arcsec \times 45 \arcsec
{\rm \mathrm cosec} |\delta|$. This catalog has 48850 sources above a limiting peak
brightness of 10 mJy beam$^{-1}$. The region surveyed is $245 \degr \leq l \leq
365 \degr$ and $|b| \leq 10 \degr$. Actually, it is the Galactic counterpart of
the SUMSS catalog.

At X-ray wavelenghts, the 2XMM catalog contains 246897 X-ray source
detections above processing likelihood threshold of 6. The median flux (in
the total photon energy band 0.2-12 keV) of the catalog detections is $\sim$
2.5$\times$10$^{-14}$ erg cm$^{-2}$ s$^{-1}$; in the soft energy band (0.2--2~keV)
the median flux is $\sim 5.8 \times 10^{-15}$  erg cm$^{-2}$ s$^{-1}$ and in the 
hard band (2--12 keV) it is $\sim 1.4 \times 10 ^{-14}$ erg cm$^{-2}$ s$^{-1}$. 
About 20\% of the sources have fluxes below 10$^{-14}$ erg cm$^{-2}$ s$^{-1}$. 
The positional accuracy of the detections in the catalog is
generally $<$ 5 arcseconds (99\% confidence radius). To date, 1.93\% of
the whole sky has been observed with the {\it XMM-Newton} satellite.

Since we are interested in finding a positional correlation between radio
and XMM-${\it Newton}$ sources, we adopted the following procedure for the determination of 
positional coincidences:

\begin{itemize}

\item First, we cross-identificated radio and X-ray sources using the method
described in the next subsection.

\item The position of those radio/XMM coincident sources were subsequently 
cross-correlated with the 2MASS\footnote{http://irsa.ipac.caltech.edu/cgi-bin/Gator/}
catalog \citep{cutri03}, using for each pair of infrared/X-ray sources the same
criteria used above.

\item Then, we inspected the SIMBAD database and the NASA/IPAC Extragalactic
Database (NED) to identify those previously known sources of the sample, which
were separated in different tables.

\item We finally studied the X-ray properties of unidentified X-ray sources with
radio and infrared counterparts and compare them with those of well-known objects
with the aim of making a preliminary diagnosis their possible nature.

\end{itemize}

\subsection{Coincidence sample and statistical analysis}

The cross-correlation of two catalogs to search for positional coincidences
of sources is usually done using, as the selection criterion, the angular distance
between sources (one in each catalog) appropriately weighted by its
uncertainty. However, as the significance of such an approach also depends on
the density of sources in the respective catalogs, the construction of the sample of
coincident sources must be carried out with some care. In this section we describe our
approach to this task, which was applied to cross-correlate the 2XMM catalog
with each one of the radio catalogs (NVSS, SUMSS, MGPS2) independently.

We constructed our sample of X--radio source coincidences in three steps.
First, we computed for each pair of sources (one in the 2XMM catalog and the
other in the corresponding radio catalog) the $R$ statistic defined as

\begin{equation}
R = \left[\frac{\left(\alpha_X - \alpha_R \right)^2}{\sigma_{\alpha,X}^2 +
\sigma_{\alpha,R}^2} +
\frac{\left(\delta_X - \delta_R \right)^2}{\sigma_{\delta,X}^2 +
\sigma_{\delta,R}^2} \right]^{1/2},
\end{equation}

\noindent
where $(\alpha_{X/R}, \delta_{X/R})$ are the equatorial coordinates of the
X-ray/radio source, and $(\sigma_{\alpha,X/R}, \sigma_{\delta,X/R})$ their
corresponding standard deviations. Clearly, $R$ increases with the increase of  
the source differential position, in such a way that the
uncertainties in the coordinates of each source are fully taken into account.
Low $R$ values point to a possible coincidence, while high $R$ values suggest
no relationship between the sources. Under the assumption that the positions of
both sources do coincide, $R$ has a Rayleigh distribution 
%(e.g., Allington-Smith et al. 1982), 
\citep[e.g.][]{allington82}, i.e. the probability that $R$ is greater than
any given non-negative value $R_0$ is

\begin{equation}
P(R > R_0) = \exp(-R^2/2).
\end{equation}

\noindent
For $R_0 = 3.03$, $P(R > R_0)$ is only of 1\%, hence we constructed a first
sample of coincident sources by retaining all pairs of sources for which $R
\leq 3.03$. For such purpose, we used our IDL (Interactive Data 
Language\footnote{http://physics.nyu.edu/grierlab/idl$_{-}$html$_{-}$help/home.html})-based 
code to cross-correlate the aforementioned catalogs. Given the
construction of this sample, the probability that a {\em true} coincidence is
missed is only 1\% (the completeness is 99\%). However, we expect this sample
to be contaminated by some amount of chance coincidences of unrelated X and
radio sources. We call it our {\em dirty} sample.

%-----------------------------------------------------------------------------
\begin{table*}
\label{sampletab}
\small
\caption{Statistical analysis.}
\begin{tabular}{lcccccccccccc}
\tableline
Catalogue & $\sigma_{\alpha,R}$ & $\sigma_{\delta,R}$ & $\langle n \rangle$
& $\sigma_n$ & $P_{\mathrm u}(R \leq 3.03)$ & $N$ & $\langle N_{99} \rangle$ &
$\sigma_{99}$ & $N_{\mathrm 99}$ & $f_{99}$ & $f_{95}$ & $f_{68}$ \\
 & $\arcsec$ & $\arcsec$ & $\deg^{-2}$ & $\deg^{-2}$ & $\times 10^{-4}$ & & & & & & \\
\tableline
MGPS2   & 2.9 & 3.1 & 19 &  6 &  5.5 &  24148 &  13 &  4 &  109 & 0.12 & 0.09 & 0.06 \\
SUMSS A & 2.7 & 2.9 & 30 &  5 &  8.0 &  17436 &  14 &  4 &  232 & 0.06 & 0.05 & 0.02 \\
SUMSS B & 2.7 & 2.9 & 20 &  5 &  5.3 &  17681 &   9 &  3 &  217 & 0.04 & 0.03 & 0.02 \\
NVSS    & 4.0 & 4.3 & 52 & 10 & 24.5 & 150032 & 368 & 19 & 2762 & 0.13 & 0.10 & 0.07 \\
\tableline
\end{tabular}
\end{table*}
%---------------------------------------------------

As a second step, we estimated the fraction of spurious (chance) coincidences
in our sample. For this purpose, we evaluated the probability $P_{\mathrm u}(R
\leq R_0)$ of getting an unrelated 2XMM--radio source pair with $R \leq R_0$,
assuming a uniform distribution of radio sources in the vicinity of each 2XMM
source. This probability is

\begin{eqnarray}
\nonumber
P_{\mathrm u}(R \leq R_0) = \int_0^\infty \int_0^\infty P_{\mathrm u}(R \leq R_0 |
\sigma_\alpha, \sigma_\delta) \cdot \\
f(\sigma_\alpha, \sigma_\delta) \mathrm{d}\sigma_\alpha {\mathrm d}\sigma_\delta,
\label{probr0}
\end{eqnarray}

\noindent
where $\sigma_\alpha = (\sigma_{\alpha,X}^2 +
\sigma_{\alpha,R}^2)^{1/2}$, $\sigma_\delta = (\sigma_{\delta,X}^2 +
\sigma_{\delta,R}^2)^{1/2}$, $P_{\mathrm u}(R \leq R_0 | \sigma_\alpha,
\sigma_\delta)$ is the conditional probability that an unrelated pair has $R
< R_0$, given the values of $\sigma_\alpha$ and $\sigma_\delta$, and
$f(\sigma_\alpha, \sigma_\delta)$ the joint probability density function of
these two variables. It can be seen, from the definition of $R$, that
$P_{\mathrm u}(R \leq R_0 | \sigma_\alpha, \sigma_\delta)$ is the probability
of finding at least one radio source inside an ellipse of semiaxes
$\sigma_\alpha$ and $\sigma_\delta$ centered at the position of the 2XMM
source. If the local density of radio sources is $n$, this is simply

\begin{equation}
P_{\mathrm u}(R \leq R_0 | \sigma_\alpha, \sigma_\delta) = 1 - {\mathrm
e}^{-\pi n R_0^2 \sigma_\alpha \sigma_\delta}.
\end{equation}

\noindent
To simplify expression~\ref{probr0}, we computed mean position uncertainties
of the radio sources directly from the data in the catalogs, obtaining values
$< 5\arcsec$ (see Table 1). Mean position uncertainties in the
2XMM catalog were also computed, obtaining a value of $2.0\arcsec$. We also 
determined the radio source density $n$ at the position of each 2XMM
source located within the radio catalog boundaries, by counting the number of
sources inside $1\degr$-radius circles centered in the X sources, and dividing
it by the area of the circles. For this purpose, we divided the SUMSS catalog
into two parts, one comprising sources with $\delta_{J2000} \leq -50\degr$ (SUMSS
A), and the other comprising sources with $\delta_{J2000} > -50\degr$ (SUMSS B)
because of the different sensitivities of this catalog in the two regions,
which results in different source densities. The mean density $\langle n
\rangle$ for each radio catalog is given in Table 1, together
with the dispersion $\sigma_n$. Given that $\sigma_n$ is small compared to
$\langle n \rangle$, for the purpose of computing the probability of chance
coincidences we assumed that the local source density $n$ in each radio
catalog is constant and equal to $\langle n \rangle$.

From the source densities and mean positional uncertainties, it is easy to
verify that in every case $\pi \langle n \rangle R_0^2 \sigma_\alpha
\sigma_\delta \ll 1$ hence, to first order, $P_{\mathrm u}(R \leq R_0 |
\sigma_\alpha, \sigma_\delta) = \pi \langle n \rangle R_0^2 \sigma_\alpha
\sigma_\delta$. Assuming that total uncertainties in each coordinate are
independent, the probability of a chance coincidence is

\begin{equation}
P_{\mathrm u}(R \leq R_0) = \pi \langle n \rangle R_0^2
\langle\sigma_\alpha\rangle \langle\sigma_\delta\rangle.
\end{equation}

\noindent
Finally, by counting the number $N$ of 2XMM sources within the boundaries of
each radio catalog, we were able to compute the mean $\langle N_{99} \rangle$
and variance $\sigma^2_{99}$ of the number of coincidences expected by chance
between each pair of catalogs, at 99\% completeness,

\begin{equation}
\langle N_{99} \rangle = N P_{\mathrm u}(R \leq 3.03),
\end{equation}

\noindent
and

\begin{equation}
\sigma^2_{99} = N P_{\mathrm u}(R \leq 3.03) (1 - P_{\mathrm u}(R \leq 3.03)),
\end{equation}

\noindent
respectively. These values, together with the actual number of coincidences
$N_{99}$ obtained in the cross-correlations, are given in
Table 1. As it can be seen, the observed $N_{99}$ values are
greater than their expected means by at least $24 \sigma_{99}$, which implies
that the observed coincidences cannot be explained as due only to chance
superposition of unrelated X and radio sources. This strongly suggests that
many 2XMM sources do emit at radio wavelengths.

The contamination fraction of our sample can be estimated as $f_{99} = \langle
N_{99} \rangle / N_{99}$, and is also listed in Table 1. The
contamination is low, reaching only 13\% in the worst case. Nevertheless, it
could be interesting to have a sample with a very low contamination, at the
expense of a smaller completeness. Starting with the value of the desired
completeness, any such sample could be constructed using the data of our dirty
sample and the formulae given above. As the last step of our coincidence
analysis, and with the purpose of obtaining cleaner samples for the subsequent
analysis of the 2XMM sources, we constructed an {\em intermediate} and a {\em
clean} sample, defined by their completeness levels of 95\% ($R_0 = 2.45$) and
68\% ($R_0 = 1.51$) respectively. The contamination fractions obtained
($f_{95}$ and $f_{68}$ respectively) are shown in Table 1.

\section{Main results and discussion}
\label{results}

\subsection{Coincidences between radio/infrared and X-ray sources}

%-----------------------------------------------------------------------------
\begin{figure*}[t!] %figure 1 \center
\resizebox{1.0\hsize}{!}{\includegraphics[angle=0]{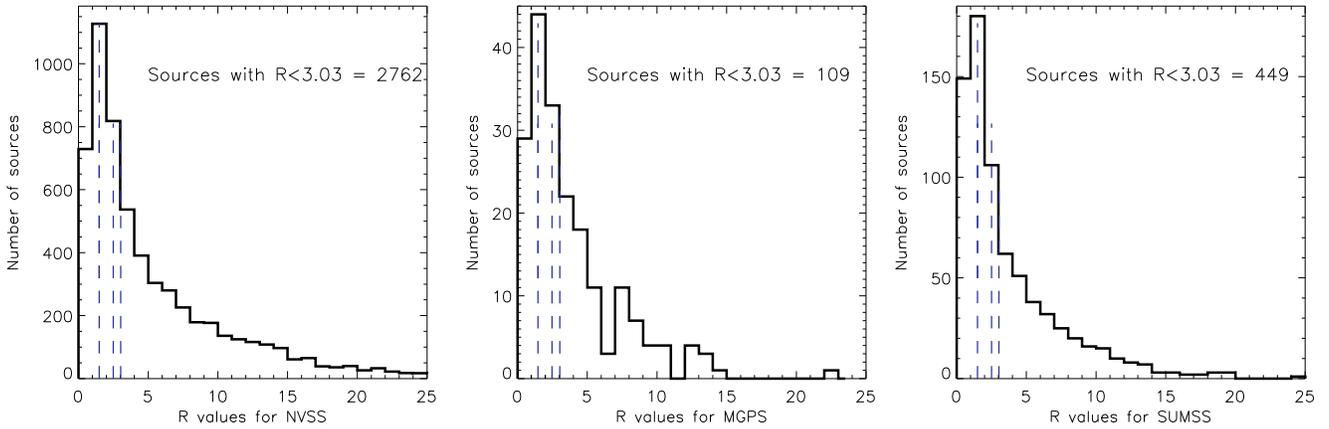}} 
\caption{Distribution of $R$ values for X-ray sources and their nearest 
neighbours. Dotted lines at $R = 3.03$, 2.45 and 1.51 define the criteria 
for the construction of our dirty, intermediate and clean samples of 
coincidences. See text.} 
\label{fig:h1} 
\end{figure*}
%-----------------------------------------------------------------------------

Fig.1 shows the distribution of $R$ for all the X-ray sources and their nearest
radio neighbours in each of the three catalogs. The
distribution shows a maximum at $R \sim 1.5$. After applying the first
criterion ($R_0 = 3.03$), a total of 2762, 449, and 109 observed coincidences
were found for the NVSS, SUMSS and MGPS2 catalogs, respectively. Among these,
only 368, 23 and 13 respectively are expected to be due to chance alignment of
unrelated sources. Figure 2 shows the all sky distribution of the positionaly correlated radio/X-ray sources.

We cross-identified the radio-X-ray sample in our 'dirty' 
sample with the 2MASS catalog using the same value of $R \leq 3.03$. Out of
2762 NVSS sources, 1253 lie at high galactic latitudes ($|b| \geq  10\degr$)
and, of them, 432 display 2MASS counterparts. At low galactic latitudes ($|b| \leq 10 \degr$) 
we found 1509 sources , of which only 468 present 2MASS
counterparts. A preliminary study of the latter sources was carried out by
Combi et al. (2008). Of the 449 SUMSS/2XMM coincidences sources, 348
are at low galactic latitude ($|b| \leq 10 \degr$), of which 17 display 2MASS
counterparts. At high galactic latitude, there are 65 coincidences, among which 27
present 2MASS counterpart. Finally, out of 109 MGPS2/2XMM coincidences (all
with $|b| \leq 10 \degr$) only 53 display 2MASS counterparts.

As a further step, we inspected the SIMBAD and the NASA/IPAC Extragalactic
Database (NED) to classify unidentified and well-known sources. In a series of
tables, described in a latter subsection, we present geometric and physical 
characteristics of different identified objects. Of the total coincident radio/X-ray
sources, 2225 are unidentified (67$\%$). These sources are published online only. 
Out of them, 416 were detected in radio catalogs different from the NVSS/SUMSS and 
MGPS2. We find a very wide taxonomy of sources
among the latter: 320 normal galaxies, 78 Seyferts I, 91 Seyferts II, 23 Liners, 
46 radiogalaxies, 12 double galaxies, 176 clusters of galaxies, 196 QSOs, 58 different 
types of stars, 15 pulsars, 4 HII regions, 19 supernovae or supernova
remnants, 5 low or high mass binaries, 2 young stellar objects, 7
young or old open clusters, 4 maser sources, 5 possible connections with
gamma-ray sources, 29 star forming regions, 4 planetary nebulae, and one source
likely related to the Galactic Center. 

%-----------------------------------------------------------------------------
\begin{figure*}[t!] %figure 2 
\resizebox{\hsize}{!}{\includegraphics[angle=0]{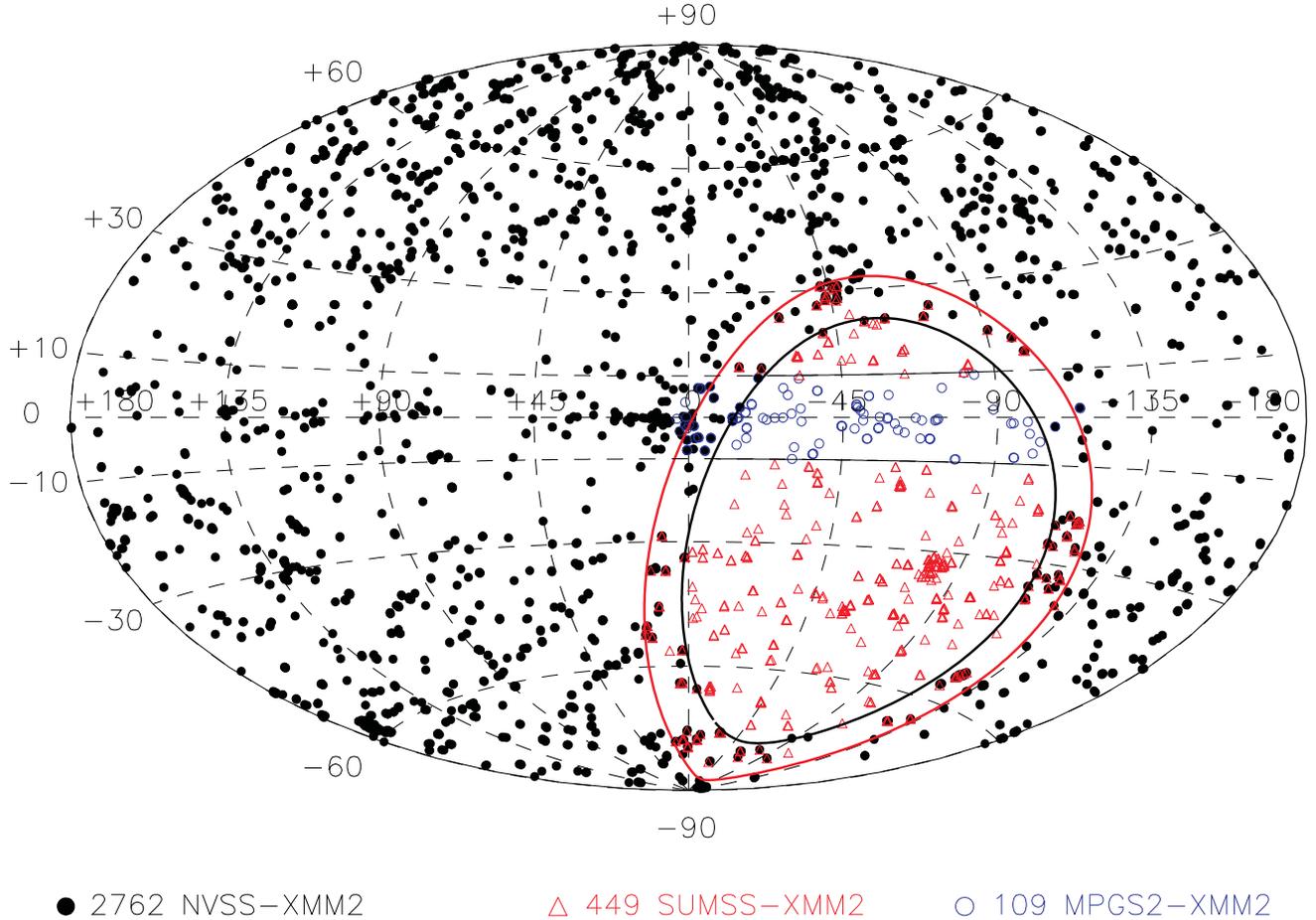}} 
\caption{Spatial distribution of the positionaly correlated radio/X-ray
sources. NVSS sources with galactic latitudes $|b|$ $\geq$ 10$^{\circ}$ 
and $|b|$ $\leq$ 10$^{\circ}$ are shown as black dots. SUMSS sources 
are shown in open blue circles and MGPS-2 sources are shown as red 
triangles. The black solid line represents the declination limit $-$40$^{\circ}$} 
\label{fig:integra} 
\end{figure*}
%-----------------------------------------------------------------------------

\subsection{X-ray properties of well-known populations} 

Statistical studies of X-ray sources based on X-ray colors can be used to classify objects with different spectral energy
distributions belonging to a class of galactic or extragalactic population. Owing to the wide energy range of the 
XMM-{\it Newton} telescope (0.2-12 keV), we were able to compute X-ray colors of sources in three different broad-bands, thus helping us to better unmask signs of high energy processes in our list of coincident sources. 
%
%Fluxes were determined for each X-ray source from its hardness-ratio. We created a grid of models with different parameters and determined %hardness ratios for them. For our study, we used both a thermal and a power-law model. Then, the 

The most efficient representation of the hardness-ratio plane in terms of thermal and/or non-thermal models suggests 
us a restricted energy range. We used three energy bands defined in the catalog as: Soft ($S$: 0.5$-$1.0 keV), Medium ($M$: 1.0$-$2.0 keV), and
Hard ($H$: 2$-$4.5 keV). The X-ray hardness-ratio was defined as: $H_{\rm x}$ = ($M-S$)/($S+M$) and $H_{\rm y}$ = ($H-M$)/($H+M$). Since it is difficult to classify individual sources with confidence on the basis of X-ray colors alone, we computed and plotted the predicted loci for thermal (in orange) and non-thermal (black) models with absorption. 
For the thermal model we used the APEC \citep{smi01} with values of temperature ranging from 0.2 to 2.6 keV and interstellar absorptions N$_{\rm H}$ from 0.1$\times$10$^{21}$ to 10$\times$10$^{22}$ cm$^{-2}$. On the other hand, for the absorbed power-law (PL) %; $N(E) \propto E^{-\Gamma}$) 
model, we used values of $\Gamma$ index ranging from 0 to 4 and interstellar absorptions N$_{\rm H}$ from 1$\times$10$^{21}$ to 4$\times$10$^{22}$ cm$^{-2}$. The grid was calculated with the Portable Interactive Multi-Mission Simulator (PIMMS6). To determine the X-ray properties of each source, we performed a two-dimensional interpolation with the grids of both models using the information provided by the hardness-ratios. First, $H_{\rm y}$ was used to determine the temperature in the thermal model and $\Gamma$ in the power-law one. Then, a value of N$_{\rm H}$ was determined for each model from $H_{\rm x}$. Our final results are given in the tables in the appendix (see Appendix).

%---------------------------------------------------
\begin{table}[!t]
\caption{Number of sources of known class
of our sample that could be fitted to each model.}
\small
\begin{tabular}{lccccc}
\tableline
%\noalign{\smallskip}
         &                                  & \multicolumn{4}{c}{Fit} \\
%\noalign{\smallskip}
\cline{3-6}
%\noalign{\smallskip}
Tipo & N$_\mathrm{tot}$ & Both & PL & Thermal & None \\  
\tableline
%\noalign{\smallskip}
Stars & 58 & 28 & 7 & 5 & 16 \\
Pulsars & 15 & 3 & 3 & 4 & 5 \\
SNe and SNR$^\mathrm{a}$ & 19 & 4 & 0 & 1 & 4 \\
X-ray binaries & 5 & 4 & 0 & 1 & 0 \\
SFR$^\mathrm{b}$ & 29 & 15 & 6 & 5 & 3 \\
HII regions & 4 & 0 & 2 & 1 & 1 \\
YSO$^\mathrm{c}$ & 2 & 2 & 0 & 0 & 0 \\
Masers & 2 & 1 & 0 & 0 & 1 \\
PNe$^\mathrm{d}$ & 4 & 0 & 1 & 1 & 2 \\
Stellar clusters & 7 & 5 & 1 & 1 & 0 \\
$\Gamma$-Ray sources & 5 & 4 & 0 & 1 & 0 \\
Galactic center & 1 & 0 & 0 & 0 & 0 \\
%\noalign{\smallskip}
\tableline
\end{tabular}
\tablenotetext{a}{Supernovae and Supernova Remnants}
\tablenotetext{b}{Star-Forming Regions}
\tablenotetext{c}{Young Stellar Objects}
\tablenotetext{d}{Planetary Nebulae}
\label{default}
\end{table}
%-----------------------------------------

The uncertainties in the estimate of the source parameters depend on the obtained values for them. For instance, N$_{\rm H}$ remains undetermined using both models for values above $4 \times 10^{22}$ cm$^{-2}$, i.e. for high $H_{\rm x}$ values ($H_{\rm x}$ $\sim 1$). Also, in the thermal model, the temperature above 2 keV is less constrained since the separation in the grid between models is very short. The X-ray characteristics for different objects are presented and briefly discussed below.

\subsubsection{Galactic sources}

The resulting hardness ratio (HR) diagrams for galactic objects are shown in Figure 3 (only the results for stars are shown here; figures for the remaining objects are available on-line). Filled (blue) points, open (green) squares and open (red) diamonds indicate the positional coincident XMM-Newton sources with NVSS, SUMSS and MGPS2 radio sources, respectively. Number of sources follows those that are presented in their respective electronic tables. Table 2 displays different types of galactic sources, which are fitted by a thermal model, a power-law model, a combination of both models or none model.

%-----------------------------------------------------------------------------
\begin{figure}[t] %figure 3a 
\includegraphics[width=8.3cm]{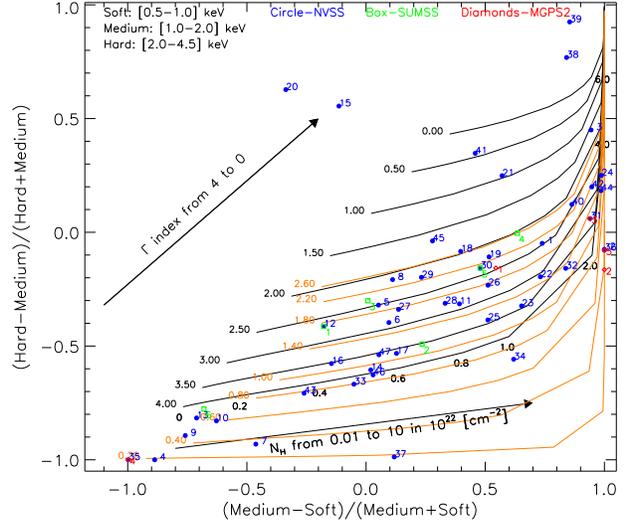} 
\caption{Ratios of source counts in different spectral 
bands for stars in the NVSS (circle), SUMS (square) and 
MGPS2 (diamonds). Plots for the remaining galactic objects
are available on-line. A color version of this plot is also 
available on-line.
} 
\label{f3} 
\end{figure}
%-----------------------------------------------------------------------------

\subsubsection{Extragalactic objects}

In the appendix, we show the HR diagrams for Quasi-Stellar Objects (QSOs) at different redshifts ($z$=0.1, 0.25, 0.5, 1.0, 1.5 and 2). 
Figures for the remaining extragalactic objects are available on-line.

{\bf Normal galaxies:} In Figure 8 we show the distribution of normal galaxies in the HR diagram. Most of these objects lie at $z$$\approx$0.1. At this redshift there are twenty six sources that do not fit the models, 56 are fitted by both models, 22 only by a PL and 24 only by a thermal model. Most of the sources lie between values of the $\Gamma$ index from 1.75 to 4, temperatures in the range 0.8 to 2.6 kT and interstellar absorption N$_{\rm H}$ with values ranging from 0.1 to 0.8$\times$10$^{22}$ cm$^{-2}$. For values of $z$ between 0.25 to 0.50 the behaviour is similar to the previous case. At values of $z$$>$1 only a PL model fit the objects. There are not objects for values of $z$$>$ 1.25.

{\bf Seyfert I:} Figure 9 shows the distribution of Seyfert I galaxies in the HR diagram. As can be seen, a large fraction of these sources have $z$$\approx$0.1, with values of $\Gamma$ indices from 2.0 to 3.75, temperatures ranging from 1.2 to 2.6 kT and a mean interstellar absorption N$_{\rm H}$ of 0.3$\times$10$^{22}$ cm$^{-2}$. It is interesting to note that when $z$ increases ($z$=0.5, $z$=1.0 and $z$=1.5) the N$_{\rm H}$ increases gradually to 0.8, 1.5 and 3 $\times$10$^{22}$ cm$^{-2}$. In addition, when $z$ increases thermal models compressed towards high values of $\Gamma$. At values of $z$$>$ 0.75 a PL model is most suitable to fit the sources. 

{\bf Seyfert II:} Figure 10 shows the distribution of Seyfert II galaxies in the HR diagram. As in the previous case, most of the sources have $z$$\approx$0.1 with values of $\Gamma$ indices from 1.5 to 4.0, temperatures ranging from 0.9 to 2.6 kT and a mean interstellar absorption N$_{\rm H}$ of 0.2 $\times$10$^{22}$ cm$^{-2}$. As can be seen, Seyfert II galaxies display values of interstellar absorption N$_{\rm H}$ lower than Seyfert I, and most of the sources are best fitted by a PL model.  

{\bf Liners:} Figure 11 shows the distribution of Liner galaxies in the HR diagram. Around 90$\%$ of the sources have $z$$\approx$0.1 and lie concentrated between values of $\Gamma$ indices of 2.25 to 3.25, temperatures ranging from 1.4 to 2.2 kT and a mean interstellar absorption N$_{\rm H}$ of 0.3 $\times$10$^{22}$ cm$^{-2}$.

{\bf Radiogalaxies:} Figure 12 shows the distribution of radiogalaxies in the HR diagram. Around 50$\%$ of the sources lie at $z$$\approx$0.1, which display values of $\Gamma$ indices from 2.25 to 4.0, temperatures ranging from 0.8 to 2.6 kT and a mean interstellar absorption N$_{\rm H}$ of 0.4 $\times$10$^{22}$ cm$^{-2}$. At values of $z$$>$ 0.5 a PL model seem to be most suitable to fit the sources. 

{\bf Double galaxies:} Figure 13 shows the distribution of double galaxies in the HR diagram. All the sources lie at $z$$\approx$0.1. Unfortunately in this case the statistic is very low, therefore, it is impossible to suggest a possible trend for these sources.

{\bf Clusters of galaxies:} Figure 14 illustrates the distribution of clusters of galaxies in the HR diagram. In this case, 89$\%$ of the objects lie at $z$$\approx$0.1 and display values of $\Gamma$ indices between 2.0 and 4.0, temperatures ranging from 0.4 to 2.6 kT and a mean interstellar absorption N$_{\rm H}$ of 0.2 $\times$10$^{22}$ cm$^{-2}$. Although most of the sources are fitted by a combination of thermal and non-thermal models, an important fraction of these (30$\%$) is only fitted by a thermal model. At values of $z$ larger than 1.0 a PL model is most suitable to fit the sources.

{\bf Quasars:} Figure 4 shows the distribution of Quasars (QSOs) in the HR diagram. Out of 196 objects, 65$\%$ lie at $z$$>$ 0.75. Objects with values of $z$  between 0.1 and 0.75 are best fitted by a combination of both thermal and PL models. On the contrary, for values of $z$$>$0.8 only a PL model fit properly the X-ray emission of the sources. In this last case, the objects show values of $\Gamma$ indices between 1.5 and 3.0, and a mean interstellar absorption N$_{\rm H}$=0.3$\times$10$^{22}$ cm$^{-2}$. 

It is interesting to note that for values of interstellar absorption N$_{\rm H}$ $>$ 10$^{22}$ cm$^{-2}$ the fraction of AGNs is quite low. A result consistent with the X-ray study recently carried out by \citet{watson09}. %Watson et al.(2009).

\subsection{Unidentified sources} 

Figures 5, 6 and 7 show the distribution of unidentified radio/X-ray sources in the HR diagram, for galactic latitudes $|b| \geq 10 \degr$, $|b| < 10 \degr$, $|b| \leq 5 \degr$ and $|b| \leq 2 \degr$, respectively. Grids with the predicted loci for thermal and non-thermal models were computed for $z$=0. Of the total unidentified sources, 392 lie out of the grids, 1048 are fitted by both models, 330 only by a power-law model and 354 only by a thermal model. 

Out of 2225 objects, 1018 have galactic latitudes $|b| \geq 10 \degr$, 1207 $|b| \leq 10 \degr$, 1166 $|b| \leq 5 \degr$ and 1124 $|b| \leq 2 \degr$. For galactic latitudes $|b| \leq 10 \degr$, the contamination of extragalactic objects is $<$ 6$\%$. In other words, 94$\%$ of sources with $|b| \leq 10 \degr$ lie mainly on the galactic plane and are probably galactic objects. 

Figure 5 (left panel) shows unidentified sources with galactic latitudes $|b| \geq 10 \degr$. As can be seen, a large concentration of sources lie between values of $\Gamma$ indices from 1.5 to 4.0 (with a mean of $\approx$ 2.75), values of temperature between 0.6 and 2.6 kT (with a mean kT of 1.6) and interstellar absorption N$_{\rm H}$ with values in the range from 0.2 to 0.8$\times$10$^{22}$ cm$^{-2}$ 
(with a mean of N$_{\rm H}$ $\approx$ 0.5$\times$10$^{22}$ cm$^{-2}$). X-ray spectral characteristics of well-known extragalactic objects (see Figures 3 and 8--14) suggest that unidentified sources with galactic latitudes $|b| \geq 10 \degr$ have mainly an extragalactic origin and could be different types of AGNs, in agreement with recent X-ray studies of sources with high galactic latitudes. These results support the widely accepted findings that the X-ray sky at middle and high galactic latitudes is largely dominated by AGNs \citep[see][]{barcons07,caccianiga07,watson09}.
%(see, Barcons et al. 2007; Caccianiga et al. 2007; Watson et al. 2008). 

Figure 6 (left panel) displays the distribution of unidentified sources with galactic latitudes $|b| \leq 10 \degr$ in the HR diagram. As can be seen, in contrast with the sources at high galactic latitudes, the distribution presents a more complex structure. There is still a small concentration of sources at low values of interstellar absorption, but with an important fraction of sources with N$_{\rm H}$ greater than 0.5$\times$10$^{22}$ cm$^{-2}$. These objects could be more absorbed sources belonging to the galactic plane and therefore with galactic origin. Comparing the positions of well-known galactic objects and unidentified sources in the HR diagrams, it is possible to provide constraints on the overall unidentified X-ray population and delineate some likely origin for these sources. For values of interstellar absorption N$_{\rm H}$ less than 0.5$\times$10$^{22}$ cm$^{-2}$, unidentified sources could be different types of stars, SNRs or a small fraction of AGNs behind the galactic plane. On the other hand, for values of interstellar absorption N$_{\rm H}$ greater than 0.5$\times$10$^{22}$ cm$^{-2}$ these objects could be HII regions, low/high mass X-ray binaries, young stellar objects, sources of masers or star forming regions. 
       
Finally, in order to distinguish more clearly the concentration of these objects in the HR diagrams, we constructed density maps for the sources at  galactic latitudes $|b|$ $\geq$ 10$^{\circ}$ and $|b|$ $\leq$ 10$^{\circ}$. Figures 5 (right panel) and 6 (right panel) show density maps for unidentified sources. The maps are 2D-histograms normalized to $1/N$, where $N$ is the total number of sources in the diagram. For the size of the bin in axis $x$ and $y$, we chose the median of the errors in HR1 and HR2, respectively. Contours were overplotted, each level corresponding to the percentage with respect to the total number of sources in the plot. 

For unidentified sources with galactic latitudes $|b|$ $\geq$ 10$^{\circ}$ (see Fig.13, right panel) the density map shows a feature of double-peak nature. Here, the contours were determined for the density map normalized to the maximum number of sources found in a bin in the histogram, to better distinguish the double-pick nature of the source sample in the density map. Comparing this figure with the HR diagrams of well-known extragalactic objects, we found that the highest peak at ($H-M$)/($H+M$)=(0.1$-$0.3) is formed by the contribution of Seyfert I and QSOs, and the lowest one at ($M-S$)/($M+S$)=($-$0.1$-$0.4) by normal galaxies. On the other hand, for unidentified sources with galactic latitudes $|b|$ $\leq$ 10$^{\circ}$ (see Fig.14, right panel), the density map shows a more complex structure, with a significant fraction of much more absorbed objects. These sources could be part of the dominant X-ray population of obscured (highly absorbed) and hard-spectrum sources, 
absent in earlier soft X-ray surveys. Possibly, the contribution of distant accreting low/high mass X-ray binaries, cataclysmic variables, RS CVn systems, and a large population of coronally active stars \citep{hands04}. %(see Hands et al. 2004).

\section{Summary} \label{summary}

In this paper we have presented the first positional cross-correlation study between radio, infrared and X-ray sources detected by XMM-{\it Newton}  telescope. Analysing radio and modern, more sensitive X-ray data, we found 3320 objects with positional coincidence between radio and X-ray catalogs.
The significance of the observed coincidences was evaluated through Monte Carlo simulations of synthetic sources following a well-tested protocol. As a result we found that the percentage of chance coincidences is less than 1\%.

Out of 3320 coincident sources, 1746 have galactic latitudes $|b| < 10^{\circ}$ and 1576 $|b| \geq 10^{\circ}$. Besides, 997 display infrared counterparts in the 2MASS catalog, 1095 are well-known objects and the remainder 2225 sources ($\sim$67$\%$) are unidentified. We found galactic and extragalactic objects among the list of well-known X-ray sources. Galactic objects such as stars, pulsars, HII regions, supernova and supernova remnants, low/high mass X-ray binaries, young stellar objects, young and old open clusters, masers, gamma-ray sources, star forming regions, planetary nebulae and a source possibly related to the galactic center. Among the extragalactic objects, we found normal galaxies, different types of AGNs (e.g., Seyfert I, Seyfert II, Liners, radiogalaxies and double galaxies), QSOs, and clusters of galaxies. 

As a second step, we carried out a dedicated study based on X-ray colors of well-known galactic and extragalactic objects. These X-ray hardness ratios were used to provide a crude representation of the X-ray spectrum and to make preliminary diagnosis of the nature of unidentified X-ray sources. X-ray spectral characteristics of well-known populations of objects suggest that unidentified sources with galactic latitudes $|b| \geq 10 \degr$ have mainly an extragalactic origin. In addittion, for galactic latitudes $|b| \leq 10 \degr$ we found that most of the unidentified sources lie at galactic latitudes $|b| \leq 2 \degr$. This latter result suggests that most of the unidentified sources found on the galactic plane might have a galactic origin. They could be different type of stars, SNRs, low/high mass X-ray binaries, young stellar objects, masers, star forming regions or part of some new population of X-ray sources. 

The present catalog, we hope, will be a very useful tool for researchers interested in both population and individual source studies. 

\acknowledgments
We thank the staff of the Osservatorio Astronomico di Palermo where part of this research was carried out. The authors acknowledge support by DGI of the Spanish Ministerio de Educaci\'on y Ciencia under grants AYA2007-68034-C03-02/-01, FEDER funds, Plan Andaluz de Investigaci\'on Desarrollo e Innovaci\'on (PAIDI) of Junta de Andaluc\'{\i}a as research group FQM-322 and the excellence fund FQM-5418. J.A.C., J.F.A.C., G.E.R and L.J.P. are researchers of CONICET. J.F.A.C was suportes by grant PICT 2007-02177 (SecyT). G.E.R. and J.A.C. were supported by grant PICT 07-00848 BID 1728/OC-AR (ANPCyT) and PIP 2010-0078 (CONICET). J.L-S acknowledges financial support by the PRICIT project S-0505/ESP-0237 (ASTROCAM) of the Comunidad Aut\'onoma de Madrid (Spain) and the Programa Nacional de Astronom\'{\i}a y Astrof\'{\i}sica of the Spanish Ministerio de Educaci\'on y Ciencia (MEC), under grants AYA2008-00695 and AYA2008-06423-C03-03. This research has also made use of the NASA/IPAC Infrared Science Archive, which is operated by the Jet Propulsion Laboratory, California Institute of Technology, under contract with the National Aeronautics and Space Administration.

\appendix

\section{Description of the tables}
 
Tables 3 to 8 list positional coincidences between the NVSS, SUMSS and MGPS2 surveys with the 2XMM-{\it Newton} catalog, 
for different types of sources. Unidentified sources are published online only. In the tables, the sources are ordered according to 
right ascension; part of the information on a source is arranged in twenty one and twenty two columns for galactic and extragalactic 
sources, respectively. These tables are intended as a useful tool for researchers interested in particular identifications.

Columns 1 and 2 provide the source number and name with rough information on its sky location, according to the conventional 
XMM-{\it Newton} source nomenclature. Columns 3 and 4 give the right ascension (RA) and declination (DEC) of the source for 
equinox J2000.0. The RA and DEC are given as [hh mm ss] and [$^\circ$ ' "], respectively. An indication of the accuracy of this 
position, in the form of equivalent (90 percent confidence level) error radii is indicated in Column 5. Columns 6 and 7 give the 
hardness ratio as defined in Sec.3.2. Columns 8 and 9 display the temperature ($kT$) and interstellar absorptions $N_{\rm H}$ obtained from the thermal model. Columns 10 and 11 list $\Gamma$ indexes and the interstellar absorptions $N_{\rm H}$ computed with the non-thermal model. The X-ray flux, in units of ergs s$^{-1}$ cm$^{-2}$, is indicated in Column 12. It was computed in the SXSSC using an energy conversion factor (ECF) in the 0.2-12 keV energy band. Columns 13, 14 and 15 give the name of the radio counterpart in the NVSS, SUMSS or MGPS2 surveys, its radio flux (in mJy) and R statistic, respectively. Columns 16, 17, 18 and 19 list the possible infrared counterpart in the 2MASS catalog, 
its magnitude Ks, the ($H-K\mathrm{s}$) and ($J-H$) differences, and the value of $R$ obtained from the XMM and 2MASS catalogs (Cutri et al. 2003, Skrutskie et al. 2006). Finally, Columns 20 and 21, provide the redshift (for extragalactic sources) and the main reference associated to each source. The references are the more recent study with the XMM-{\it Newton} satellite or another X-ray observatory. If the source has not been 
previously studied with any X-ray observatory, the reference indicates the detection at other wavelengths.  

\newpage

%-----------------------------------------------------------------------------
\begin{figure}[t] %figure A1 (QSOs) 
\centering
\includegraphics[width=16cm]{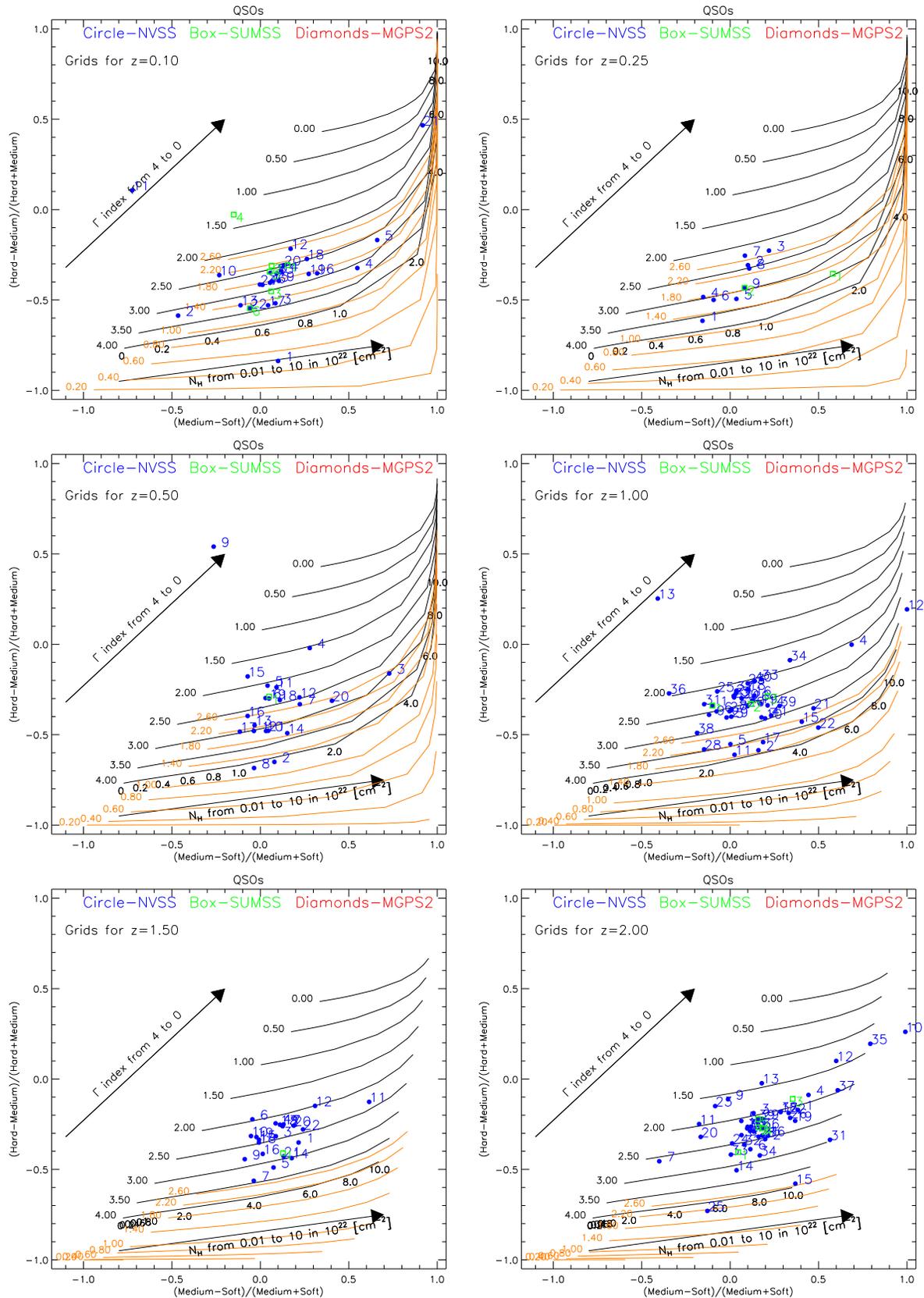} 
\caption{Ratios of source counts in different spectral bands 
for QSOs at different redshifts (z=0.1, 0.25, 0.5, 1.0, 1.5, and
2) in the NVSS (circle), SUMS (square) and 
MGPS2 (diamonds). Plots for the remaining extragalactic 
objects are available on-line. A color version of this plot is also 
available on-line.
} 
\label{fA1} 
\end{figure}
%-----------------------------------------------------------------------------

%-----------------------------------------------------------------------------
\begin{figure}[t] %figure A2 (unknown b>10) 
\centering
\includegraphics[width=8.3cm]{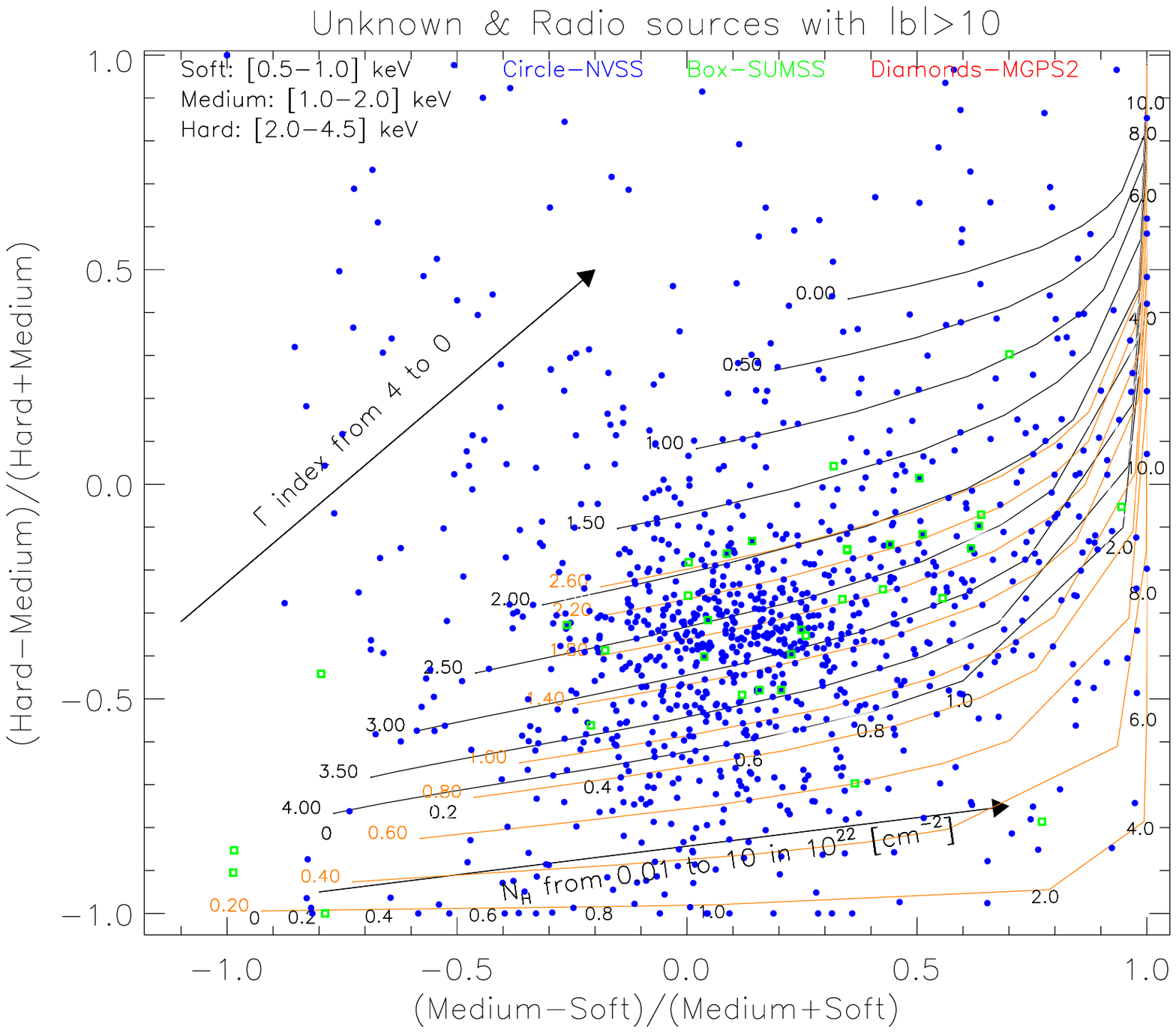}
\includegraphics[width=8.3cm]{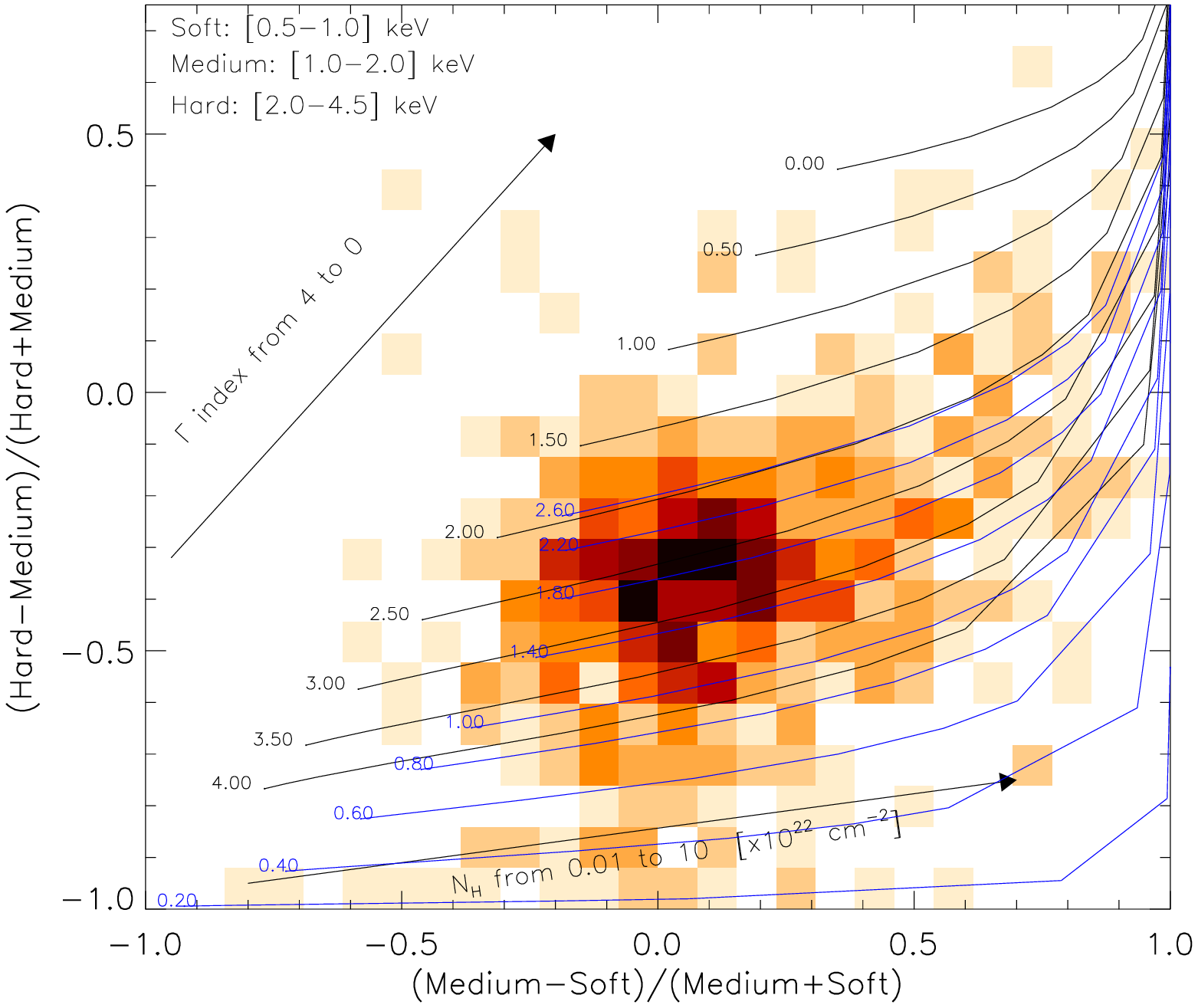} 
\caption{\textbf{Left:}
Ratios of source counts in different spectral bands for unidentified coincident radio/X-ray sources with
galactic latitude $\abs{b} > 10^\circ$. Points represent the same as in the previous figures. 
\textbf{Right:} Density map in the hardness-ratio diagram for unidentified sources with galactic latitude
$\abs{b} > 10^\circ$. Contours were overplotted, each level correspondig to the
percentage with respect to the total number of sources in the plot. Bin size correponds to a 
maximum of 39 sources per bin.
} 
\label{fA2} 
\end{figure}
%-----------------------------------------------------------------------------

%-----------------------------------------------------------------------------
\begin{figure}[t] %figure A3 (unknown b<10) 
\centering
\includegraphics[width=8.3cm]{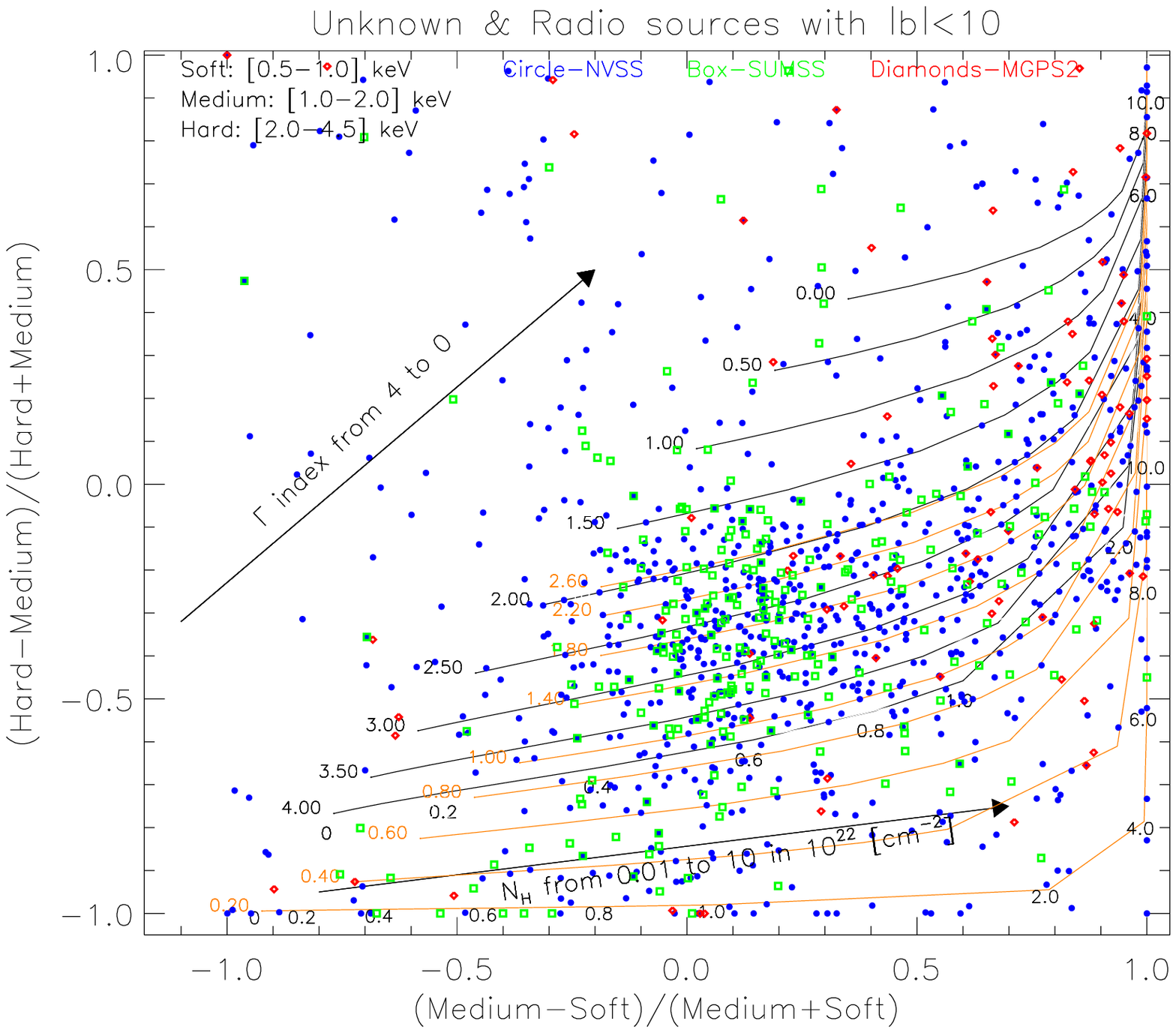}
\includegraphics[width=8.3cm]{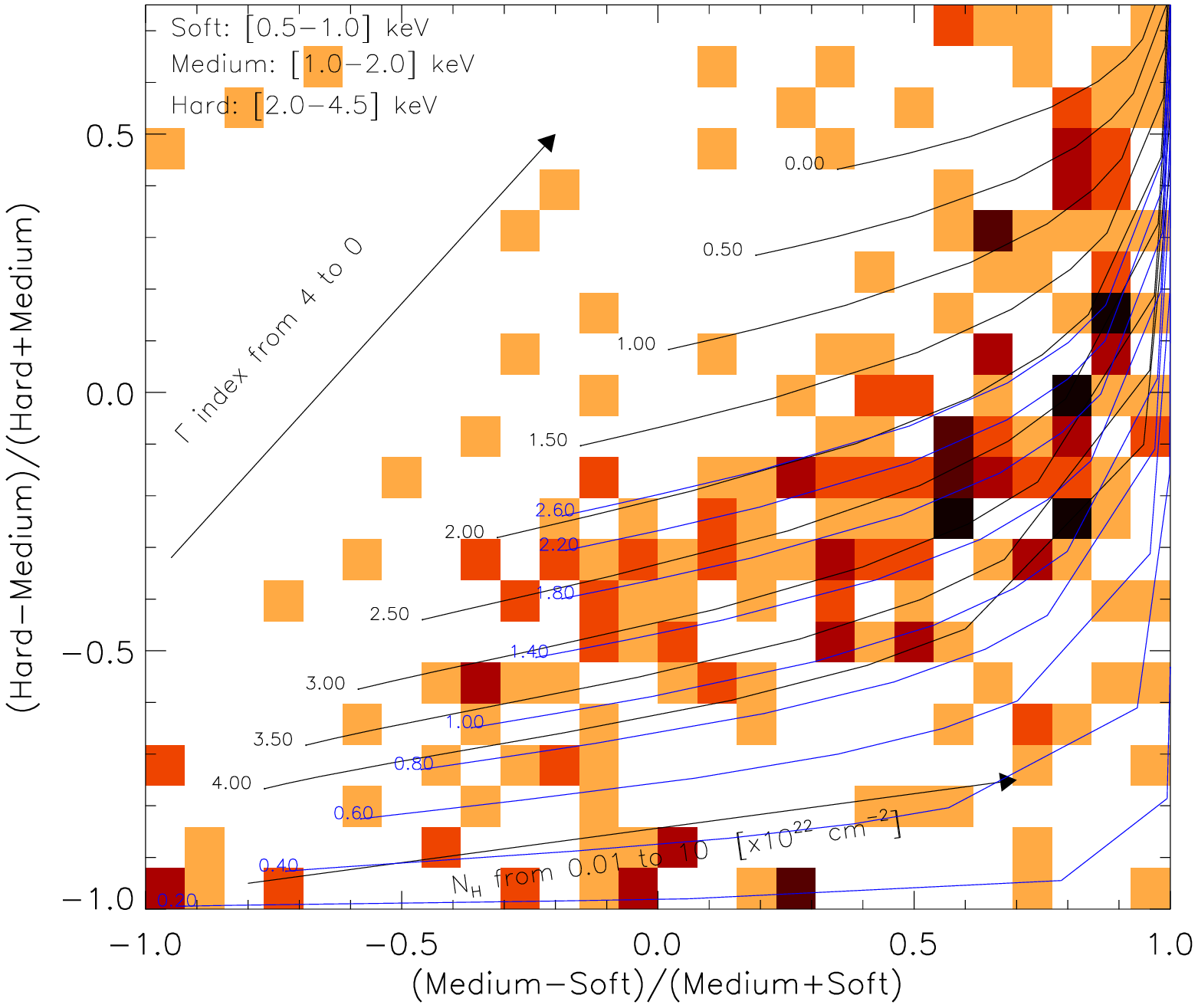} 
\caption{\textbf{Left:}
Ratios of source counts in different spectral bands for unidentified coincident radio/X-ray sources with
galactic latitude $\abs{b} < 10^\circ$. Points represent the same as in the previous figures. 
\textbf{Right:} Density map in the hardness-ratio diagram for unidentified sources with galactic latitude
$\abs{b} < 10^\circ$. Contours were overplotted, each level correspondig to the
percentage with respect to the total number of sources in the plot. Bin size correponds to a 
maximum of 39 sources per bin.
} 
\label{fA3} 
\end{figure}
%-----------------------------------------------------------------------------

%-----------------------------------------------------------------------------
\begin{figure}[t] %figure A4 (unknown b<5 and b<2) 
\centering
\includegraphics[width=8.3cm]{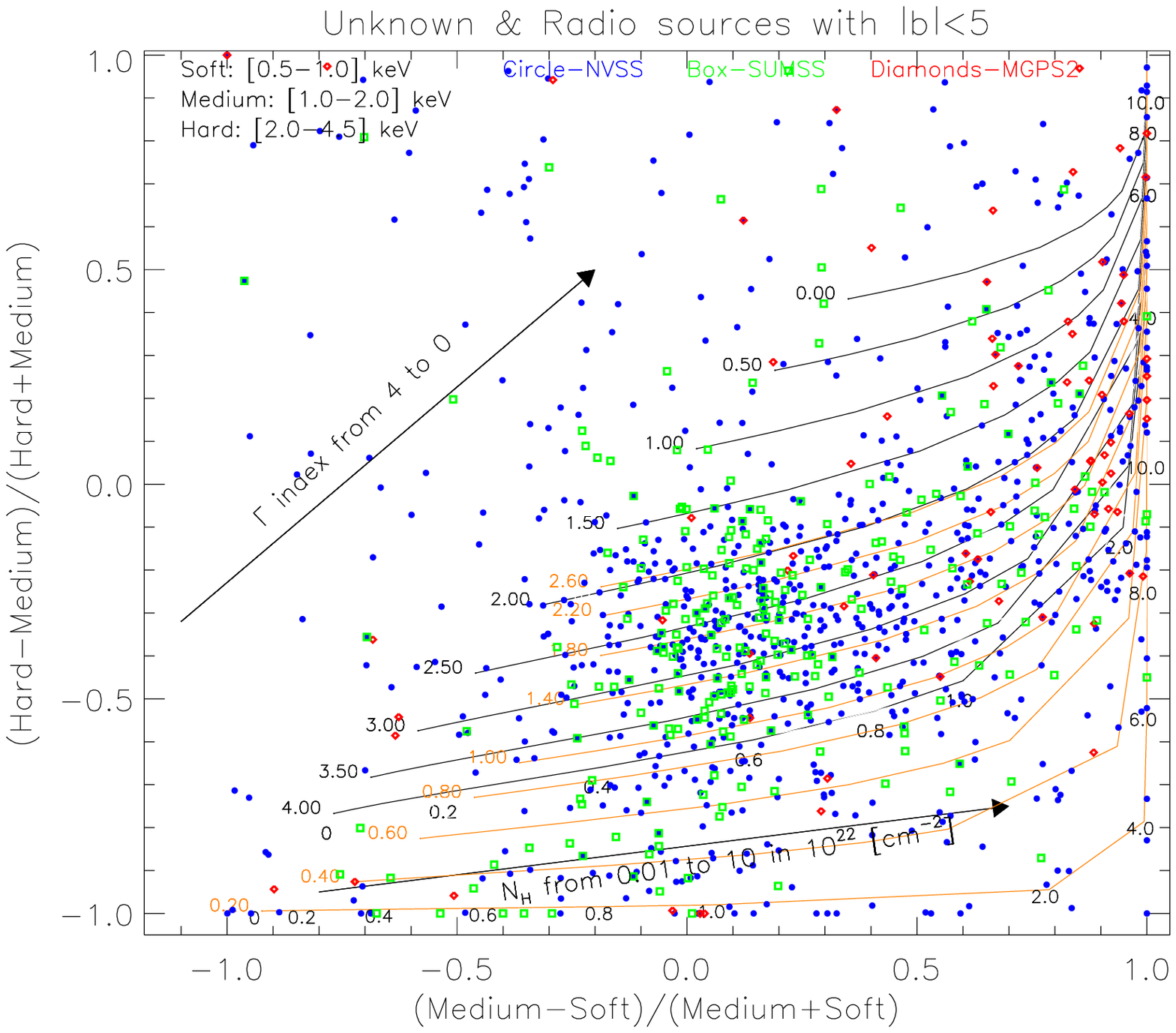}
\includegraphics[width=8.3cm]{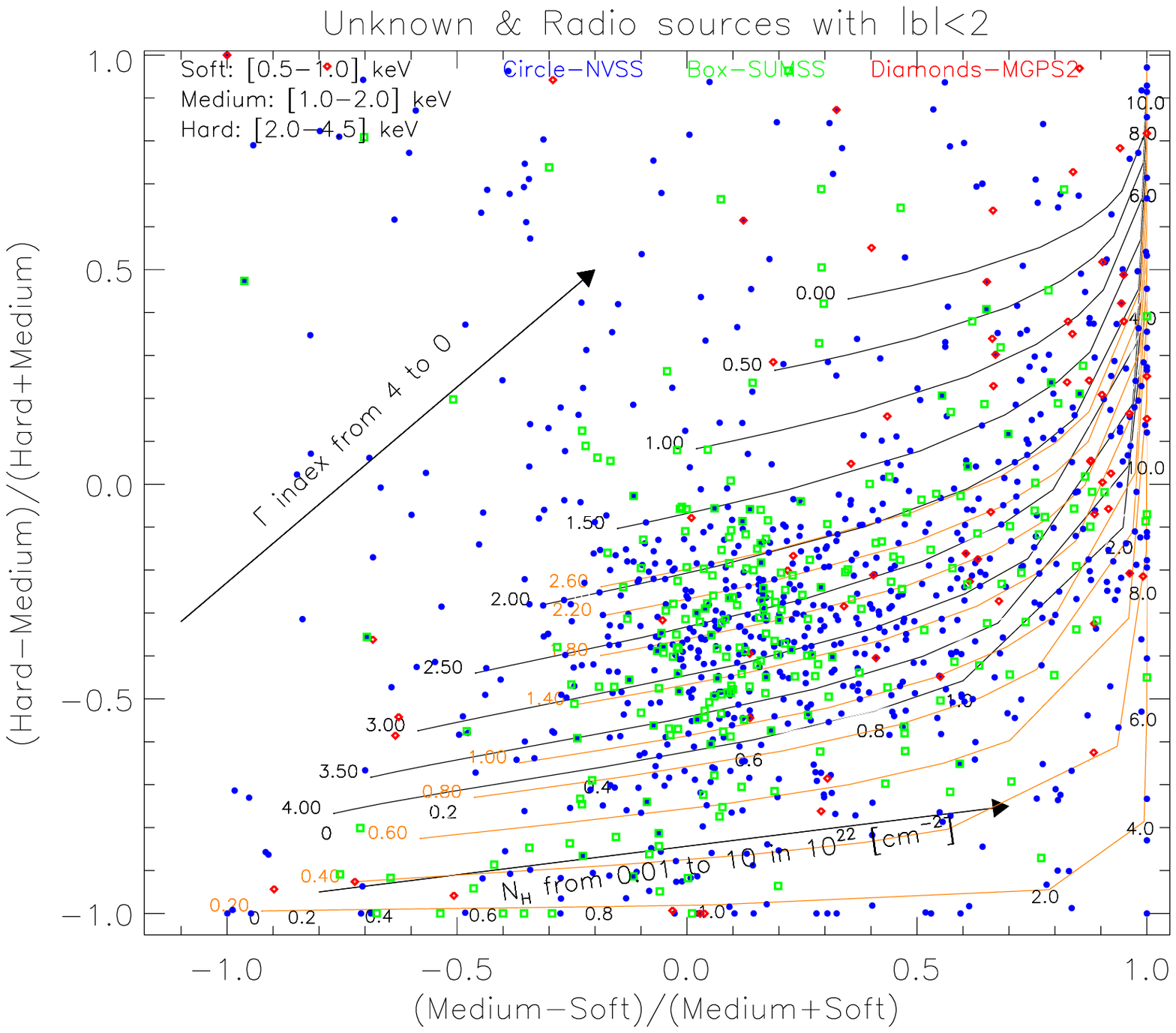} 
\caption{\textbf{Left:} Ratios of source counts in different spectral bands for 
unidentified radio/X-ray sources with galactic latitude $\abs{b} < 5^\circ$. 
\textbf{Right:} Ratios of source counts in different spectral bands for 
unidentified radio/X-ray sources with galactic latitude $\abs{b} < 2^\circ$. 
Points represent the same as in the previous figures.} 
\label{fA4} 
\end{figure}
%-----------------------------------------------------------------------------

%\include{online-figs}

%\end{document}

\newpage

%------------------------unidentified---------------------------
%\pdfpagewidth 8.5in
%\pdfpageheight 11in
%\textwidth=18.8 cm
%\textheight=25 cm
\voffset 0.30cm
%\hoffset -0.8cm
%\oddsidemargin=-1.cm
%\evensidemargin=-1.cm
%\topmargin=-1 cm
\renewcommand{\tiny}{\fontsize{2.0}{7}\selectfont}

\newpage
\begin{deluxetable}{lllllllllllllllllllll}
\tabletypesize{\tiny}
\rotate
\tablecaption{NVSS/XMM: Galactic objects} 
\tablewidth{0pt}
\tablecolumns{2}
\tablehead{
\colhead{Src.} & 
\colhead{Name} &
\colhead{RA} &
\colhead{DEC} &
\colhead{err} &
\multicolumn{2}{c}{Hardness ratio} &
\multicolumn{4}{c}{Thermal -- Non thermal} &
\colhead{F$_{\rm x}$} &
\colhead{Name} &
\colhead{F$_{\rm r}$} &
\colhead{R$_{\rm xr}$} &
\colhead{Name} &
\multicolumn{3}{c}{2MASS Mag \& Colors} &
\colhead{R$_{\rm xi}$} &
\colhead{Bib.} \\ 
\cline{8-11} \cline{17-19} 
\colhead{\#} & 
\colhead{2XMM J+} &
\colhead{h:m:s} &
\colhead{d:m:s} &
\colhead{''} &
\colhead{${{M-S}\over{S+M}}$} &
\colhead{${{H-M}\over{H+M}}$} &
\colhead{N$_{\rm H}$} &
\colhead{KT} &
\colhead{N$_{\rm H}$} &
\colhead{$\Gamma$} &
\colhead{[cgs]} &
\colhead{NVSS J+} &
\colhead{[mJy]} &
\colhead{f$_{<99}$\%} &
\colhead{2MASS J+} &
\colhead{$K$} &
\colhead{$H-K$} &
\colhead{$J-H$} &
\colhead{f$_{<99}$\%} &
\colhead{Ref.}
}
\startdata
\hline
\multicolumn{21}{c}{\bf Stars} \\
\hline
   1&  000537.0+634027 &00:05:37.00 &+63:40:27.79 & 0.45 & 0.74$\pm$0.06  &-0.05$\pm$0.05 &      0.8 &      2.0 &       1. &      2.5 &   1.25$\pm$ 0.10 & 000537+634030 &   41.4 $\pm$   1.3 &2.81 &      $------$ & $--$ & $--$ & $--$ &$--$ &     302 \\
   2&  000647.0+633602 &00:06:47.04 &+63:36: 2.37 & 1.49 & 1.00$\pm$0.40  &-0.08$\pm$0.26 &     $--$ &     $--$ &     $--$ &     $--$ &   0.08$\pm$ 0.04 & 000647+633558 &    4.5 $\pm$   0.5 &1.06 &      $------$ & $--$ & $--$ & $--$ &$--$ &     302 \\
   3&  000701.5+634152 &00:07: 1.57 &+63:41:52.19 & 0.33 & 0.94$\pm$0.04  & 0.45$\pm$0.04 &     $--$ &     $--$ &       2. &      1.5 &   2.21$\pm$ 0.13 & 000701+634141 &    3.2 $\pm$   0.5 &1.45 &      $------$ & $--$ & $--$ & $--$ &$--$ &     302 \\
   4&  000937.0-321630 &00:09:37.01 &-32:16:30.97 &34.67 &-0.89$\pm$0.12  &-1.00$\pm$0.41 &     $--$ &     $--$ &     $--$ &     $--$ &   1.15$\pm$ 0.65 & 000935-321636 &  388.3 $\pm$  11.7 &0.49 & 000937-321630 &15.46 &-0.06 &0.587 &0.21 &     474 \\
   5&  003300.9+851618 &00:33: 0.99 &+85:16:17.09 & 0.77 & 0.05$\pm$0.03  &-0.32$\pm$0.03 &      0.1 &      2.0 &      0.2 &      2.5 &   5.31$\pm$ 0.28 & 003259+851616 &   41.4 $\pm$   2.0 &2.55 &      $------$ & $--$ & $--$ & $--$ &$--$ &      28 \\
   6&  004056.6+851823 &00:40:56.62 &+85:18:23.64 & 0.73 & 0.10$\pm$0.11  &-0.40$\pm$0.13 &      0.2 &      1.6 &      0.4 &      3.0 &   0.23$\pm$ 0.05 & 004055+851825 &    8.3 $\pm$   0.5 &1.78 &      $------$ & $--$ & $--$ & $--$ &$--$ &     161 \\
   7&  013335.8+303627 &01:33:35.63 &+30:36:27.79 & 1.93 &-0.46$\pm$0.05  &-0.93$\pm$0.06 &      0.4 &     0.30 &     $--$ &     $--$ &   0.15$\pm$ 0.02 & 013335+303629 &    3.3 $\pm$   0.5 &0.44 &      $------$ & $--$ & $--$ & $--$ &$--$ &     279 \\
\enddata
\end{deluxetable}

%--------------------------normal galaxies--------------------
\newpage
\begin{deluxetable}{llllllllllllllllllllll}
\tabletypesize{\tiny}
\rotate
\tablecaption{NVSS/XMM: Extragalactic objects} 
\tablewidth{0pt}
\tablecolumns{2}
\tablehead{
\colhead{Src.} & 
\colhead{Name} &
\colhead{RA} &
\colhead{DEC} &
\colhead{err} &
\multicolumn{2}{c}{Hardness ratio} &
\multicolumn{4}{c}{Thermal -- Non thermal} &
\colhead{F$_{\rm x}$} &
\colhead{Name} &
\colhead{F$_{\rm r}$} &
\colhead{R$_{\rm xr}$} &
\colhead{Name} &
\multicolumn{3}{c}{2MASS Mag \& Colors} &
\colhead{R$_{\rm xi}$} &
\colhead{Redshift} &
\colhead{Bib.} \\ 
%}
\cline{8-11} \cline{17-19} 
%\tablehead{
\colhead{\#} & 
\colhead{2XMM J+} &
\colhead{h:m:s} &
\colhead{d:m:s} &
\colhead{''} &
\colhead{${{M-S}\over{S+M}}$} &
\colhead{${{H-M}\over{H+M}}$} &
\colhead{N$_{\rm H}$} &
\colhead{KT} &
\colhead{N$_{\rm H}$} &
\colhead{$\Gamma$} &
\colhead{[cgs]} &
\colhead{NVSS J+} &
\colhead{[mJy]} &
\colhead{f$_{<99}$\%} &
\colhead{2MASS J+} &
\colhead{$K$} &
\colhead{$H-K$} &
\colhead{$J-H$} &
\colhead{f$_{<99}$\%} &
\colhead{$z$} &
\colhead{Ref.}
}
\startdata
\hline
\multicolumn{22}{c}{\bf Normal Galaxies} \\
\hline
   1&  001110.4-120114 &00:11:10.48 &-12:01:14.43 & 0.95 &-0.35$\pm$0.10 &-0.59$\pm$0.17 &     $--$ &     $--$ &      0.2 &      3.2 &   0.36$\pm$ 0.10 & 001110-120113&   12.8$\pm$   1.0 &0.40 & 001110-120114 &12.87 &0.437 &0.692 &0.356 &0.020096&     200 \\
   2&  001501.6-390537 &00:15: 1.69 &-39:05:37.30 & 1.87 & 0.57$\pm$0.30 & 0.13$\pm$0.25 &     $--$ &     $--$ &      0.6 &      1.2 &   0.17$\pm$ 0.11 & 001501-390541&    2.8$\pm$   0.5 &1.00 & 001501-390537 &14.90 &0.362 &0.893 &2.629 &0.160000&     266 \\
   3&  001537.0-390946 &00:15:37.17 &-39:09:47.41 & 2.32 &-0.29$\pm$0.15 &-0.62$\pm$0.22 &     0.01 &      1.2 &      0.2 &      3.5 &   0.07$\pm$ 0.04 & 001537-390940&    4.2$\pm$   0.5 &2.03 & 001537-390947 &13.40 &0.588 &0.821 &0.787 &0.062843&     404 \\
   4&  002155.3-151738 &00:21:55.38 &-15:17:38.34 & 1.14 &-0.24$\pm$0.13 &-0.38$\pm$0.21 &     $--$ &     $--$ &      0.1 &      2.5 &   0.10$\pm$ 0.04 & 002155-151736&    7.7$\pm$   0.6 &1.10 & 002155-151738 &13.13 &0.528 &0.734 &1.138 &0.027476&     251 \\
   5&  002914.1+345632 &00:29:14.16 &+34:56:32.34 & 0.29 & 0.64$\pm$0.04 &-0.04$\pm$0.04 &     $--$ &     $--$ &       2. &      2.2 &   3.13$\pm$ 0.17 & 002914+345632& 1902.9$\pm$  57.1 &0.66 &      $------$ & $--$ & $--$ & $--$ & $--$ &0.517000&     179 \\
   6&  003017.4+261139 &00:30:17.49 &+26:11:39.50 & 1.88 & 0.29$\pm$0.21 &-0.94$\pm$0.19 &       1. &     0.30 &     $--$ &     $--$ &   0.06$\pm$ 0.05 & 003017+261144&    2.7$\pm$   0.4 &3.00 & 003017+261139 &13.65 &0.439 &0.775 &0.587 &0.077000&      78 \\
   7&  003432.3-054330 &00:34:32.38 &-05:43:30.65 & 1.24 &-0.36$\pm$0.16 &-0.86$\pm$0.27 &      0.2 &     0.50 &     $--$ &     $--$ &   0.24$\pm$ 0.14 & 003432-054326&    2.2$\pm$   0.5 &0.37 &      $------$ & $--$ & $--$ & $--$ & $--$ &0.013774&     428 \\
\enddata
\end{deluxetable}

\begin{deluxetable}{llllllllllllllllllllll}
\tabletypesize{\tiny}
\rotate
\tablecaption{SUMSS/XMM: Galactic objects} 
\tablewidth{0pt}
\tablecolumns{2}
\tablehead{
\colhead{Src.} & 
\colhead{Name} &
\colhead{RA} &
\colhead{DEC} &
\colhead{err} &
\multicolumn{2}{c}{Hardness ratio} &
\multicolumn{4}{c}{Thermal -- Non thermal} &
\colhead{F$_{\rm x}$} &
\colhead{Name} &
\colhead{S$_{\rm t}$} &
\colhead{S$_{\rm p}$} &
\colhead{R$_{\rm xr}$} &
\colhead{Name} &
\multicolumn{3}{c}{2MASS Mag \& Colors} &
\colhead{R$_{\rm xi}$} &
\colhead{Bib.} \\ 
\cline{8-11} \cline{18-20} 
\colhead{\#} & 
\colhead{2XMM J+} &
\colhead{h:m:s} &
\colhead{d:m:s} &
\colhead{''} &
\colhead{${{M-S}\over{S+M}}$} &
\colhead{${{H-M}\over{H+M}}$} &
\colhead{N$_{\rm H}$} &
\colhead{KT} &
\colhead{N$_{\rm H}$} &
\colhead{$\Gamma$} &
\colhead{[cgs]} &
\colhead{SUMMS J+} &
\colhead{[mJy]} &
\colhead{[mJy]} &
\colhead{f$_{<99}$\%} &
\colhead{2MASS J+} &
\colhead{$K$} &
\colhead{$H-K$} &
\colhead{$J-H$} &
\colhead{f$_{<99}$\%} &
\colhead{Ref.}
}
\startdata
\hline
\multicolumn{22}{c}{\bf Stars} \\
\hline
   1&  023917.9-341654 &02:39:17.93 &-34:16:54.02 & 2.20 &-0.18$\pm$0.08  &-0.41$\pm$0.12 &       0. &      1.8 &      0.2 &      2.7 &   0.54$\pm$  0.21 & 023917-341654 &    1.7     &   23.3 & 1.78 & 023917-341654 &15.72 &0.163 &0.540 &2.09 &     416 \\
   2&  045818.0-751637 &04:58:18.03 &-75:16:37.90 & 0.03 & 0.24$\pm$0.00  &-0.49$\pm$0.00 &      0.4 &      1.2 &      0.6 &      3.5 & 134.56$\pm$  0.44 & 045818-751637 &    1.4     &    8.8 & 1.33 & 045818-751637 &5.287 &0.194 &0.557 &1.27 &     433 \\
   3&  111058.4-761759 &11:10:58.45 &-76:17:59.26 & 1.26 & 0.01$\pm$0.07  &-0.30$\pm$0.09 &      0.1 &      2.0 &      0.2 &      2.5 &   0.23$\pm$  0.04 & 111058-761759 &    3.7     &  113.7 & 2.89 & 111058-761759 &14.44 &0.792 &0.922 &1.02 &      64 \\
   4&  125701.0-764832 &12:57: 1.01 &-76:48:32.77 & 1.52 & 0.63$\pm$0.17  &-0.00$\pm$0.18 &      0.6 &      2.6 &      0.6 &      2.0 &   0.12$\pm$  0.05 & 125701-764832 &    1.2     &    7.8 & 2.30 &      $------$ & $--$ & $--$ & $--$ &$--$ &     346 \\
   5&  155702.2-380219 &15:57: 2.25 &-38:02:19.39 & 0.65 & 0.48$\pm$0.06  &-0.16$\pm$0.06 &      0.4 &      2.2 &      0.6 &      2.2 &   3.32$\pm$  0.37 & 155702-380219 &    2.2     &   64.3 & 0.98 &      $------$ & $--$ & $--$ & $--$ &$--$ &     236 \\
   6&  201702.3-704517 &20:17: 2.34 &-70:45:17.10 & 4.98 &-0.68$\pm$0.14  &-0.78$\pm$0.50 &     $--$ &     $--$ &     $--$ &     $--$ &   0.39$\pm$  0.26 & 201702-704517 &    1.3     &   13.9 & 2.69 &      $------$ & $--$ & $--$ & $--$ &$--$ &      20 \\
\hline
\multicolumn{22}{c}{\bf Pulsars} \\
\enddata
\end{deluxetable}

\newpage
\begin{deluxetable}{lllllllllllllllllllllll}
\tabletypesize{\tiny}
\rotate
\tablecaption{SUMSS/XMM: Extragalactic objects} 
\tablewidth{0pt}
\tablecolumns{2}
\tablehead{
\colhead{Src.} & 
\colhead{Name} &
\colhead{RA} &
\colhead{DEC} &
\colhead{err} &
\multicolumn{2}{c}{Hardness ratio} &
\multicolumn{4}{c}{Thermal -- Non thermal} &
\colhead{F$_{\rm x}$} &
\colhead{Name} &
\colhead{S$_{\rm t}$} &
\colhead{S$_{\rm p}$} &
\colhead{R$_{\rm xr}$} &
\colhead{Name} &
\multicolumn{3}{c}{2MASS Mag \& Colors} &
\colhead{R$_{\rm xi}$} &
\colhead{Redshift} &
\colhead{Bib.} \\ 
%}
\cline{8-11} \cline{18-20} 
%\tablehead{
\colhead{\#} & 
\colhead{2XMM J+} &
\colhead{h:m:s} &
\colhead{d:m:s} &
\colhead{''} &
\colhead{${{M-S}\over{S+M}}$} &
\colhead{${{H-M}\over{H+M}}$} &
\colhead{N$_{\rm H}$} &
\colhead{KT} &
\colhead{N$_{\rm H}$} &
\colhead{$\Gamma$} &
\colhead{[cgs]} &
\colhead{SUMMS J+} &
\colhead{[mJy]} &
\colhead{[mJy]} &
\colhead{f$_{<99}$\%} &
\colhead{2MASS J+} &
\colhead{$K$} &
\colhead{$H-K$} &
\colhead{$J-H$} &
\colhead{f$_{<99}$\%} &
\colhead{$z$} &
\colhead{Ref.}
}
\startdata
\hline
\multicolumn{22}{c}{\bf Normal galaxies} \\
\hline
   1&  000935.5-321636 &00:09:35.58 &-32:16:36.93 & 0.11 &-0.27$\pm$0.01 &-0.48$\pm$0.02 &      0.4 &      1.0 &      0.6 &      3.7 &   7.21$\pm$  0.15 & 000935-321636&   13.7 &  453.6 &1.36 & 000935-321636 &11.55 &0.384 &0.747 &1.008 &0.025628&     410 \\
   2&  002124.5-483954 &00:21:24.53 &-48:39:54.85 & 0.39 &-0.40$\pm$0.04 &-0.69$\pm$0.07 &      0.1 &      1.6 &      0.4 &      3.0 &   0.59$\pm$  0.06 & 002124-483954&    1.5 &   21.8 &1.18 &      $------$ & $--$ & $--$ & $--$ & $--$ &0.012838&     410 \\
   3&  002131.7-483729 &00:21:31.70 &-48:37:29.11 & 0.17 & 0.13$\pm$0.02 &-0.44$\pm$0.02 &     $--$ &     $--$ &     $--$ &     $--$ &   2.69$\pm$  0.09 & 002131-483729&    7.4 &  243.8 &0.45 & 002131-483729 &10.57 &0.837 &0.971 &1.514 &   $---$&     410 \\
   4&  002238.2-483444 &00:22:38.25 &-48:34:44.21 & 2.25 &-0.23$\pm$0.17 &-0.65$\pm$0.27 &     $--$ &     $--$ &     $--$ &     $--$ &   0.12$\pm$  0.10 & 002238-483444&    1.0 &   10.8 &2.08 &      $------$ & $--$ & $--$ & $--$ & $--$ &   $---$&     410 \\
   5&  003652.3-333318 &00:36:52.32 &-33:33:18.34 & 0.45 & 0.03$\pm$0.06 &-0.44$\pm$0.07 &     $--$ &     $--$ &       2. &      1.5 &   2.21$\pm$  0.39 & 003652-333318&    2.1 &   33.4 &1.94 &      $------$ & $--$ & $--$ & $--$ & $--$ &   $---$&     171 \\
   6&  005353.2-374026 &00:53:53.30 &-37:40:24.15 & 1.43 &-0.01$\pm$0.10 &-0.12$\pm$0.11 &      0.6 &      1.0 &      0.8 &      3.7 &   0.15$\pm$  0.04 & 005353-374024&    6.7 &  215.4 &0.55 &      $------$ & $--$ & $--$ & $--$ & $--$ &   $---$&     367 \\
   7&  005403.5-374640 &00:54: 3.52 &-37:46:40.04 & 2.16 &-0.55$\pm$0.31 &-1.00$\pm$0.38 &     $--$ &     $--$ &     $--$ &     $--$ &   0.19$\pm$  0.19 & 005403-374640&    8.9 &  290.6 &0.77 &      $------$ & $--$ & $--$ & $--$ & $--$ &   $---$&     331 \\
\enddata
\end{deluxetable}

\begin{deluxetable}{llllllllllllllllllllll}
\tabletypesize{\tiny}
\rotate
\tablecaption{MGPS2/XMM: Galactic objects} 
\tablewidth{0pt}
\tablecolumns{2}
\tablehead{
\colhead{Src.} & 
\colhead{Name} &
\colhead{RA} &
\colhead{DEC} &
\colhead{err} &
\multicolumn{2}{c}{Hardness ratio} &
\multicolumn{4}{c}{Thermal -- Non thermal} &
\colhead{F$_{\rm x}$} &
\colhead{Name} &
\colhead{F$_{\rm r}$} &
\colhead{R$_{\rm xr}$} &
\colhead{Name} &
\multicolumn{3}{c}{2MASS Mag \& Colors} &
\colhead{R$_{\rm xi}$} &
\colhead{Bib.} \\ 
%}
\cline{8-11} \cline{17-19} 
%\tablehead{
\colhead{\#} & 
\colhead{2XMM J+} &
\colhead{h:m:s} &
\colhead{d:m:s} &
\colhead{''} &
\colhead{${{M-S}\over{S+M}}$} &
\colhead{${{H-M}\over{H+M}}$} &
\colhead{N$_{\rm H}$} &
\colhead{KT} &
\colhead{N$_{\rm H}$} &
\colhead{$\Gamma$} &
\colhead{[cgs]} &
\colhead{MGPS J+} &
\colhead{[mJy]} &
\colhead{f$_{<99}$\%} &
\colhead{2MASS J+} &
\colhead{$K$} &
\colhead{$H-K$} &
\colhead{$J-H$} &
\colhead{f$_{<99}$\%} &
\colhead{Ref.}
}
\startdata
\hline
\multicolumn{21}{c}{\bf Stars} \\
\hline
   1&  110728.5-653126 &11:07:28.54 &-65:31:26.38 & 1.17 & 0.54$\pm$0.13  &-0.16$\pm$0.13 &      0.40 &      2.00 &      0.60 &      2.50 &   0.56$\pm$ 0.12 & 110727-653125 &   15.6 $\pm$   1.2 &1.86 & 110728-653126 &14.83 &0.113 &0.411 &1.04 &     165 \\
   2&  134648.9-625237 &13:46:48.95 &-62:52:37.42 & 7.44 & 1.00$\pm$0.03  &-0.16$\pm$0.15 &      $--$ &      $--$ &      $--$ &      $--$ &   1.66$\pm$ 0.47 & 134647-625233 &   24.2 $\pm$   2.6 &1.50 & 134648-625237 &10.58 &0.873 &2.012 &0.75 &     360 \\
   3&  160833.8-390009 &16:08:33.89 &-39:00: 9.95 & 1.23 & 0.94$\pm$0.13  & 0.06$\pm$0.15 &       2.0 &      1.20 &       2.0 &      3.25 &   0.34$\pm$ 0.08 & 160833-390016 &   31.0 $\pm$   2.1 &2.94 &      $------$ & $--$ & $--$ & $--$ &$--$ &     162 \\
   4&  174948.2-370129 &17:49:48.21 &-37:01:29.20 & 1.86 &-1.00$\pm$0.19  &-1.00$\pm$**** &      $--$ &      $--$ &      $--$ &      $--$ &   0.08$\pm$ 0.05 & 174947-370135 &   16.0 $\pm$   1.9 &1.85 &      $------$ & $--$ & $--$ & $--$ &$--$ &     226 \\
   5&  175015.5-370616 &17:50:15.57 &-37:06:16.59 & 1.36 & 1.00$\pm$0.16  &-0.08$\pm$0.18 &      $--$ &      $--$ &      $--$ &      $--$ &   0.47$\pm$ 0.18 & 175015-370615 &  146.1 $\pm$   4.6 &0.85 & 175015-370616 &13.44 &0.137 &0.621 &2.81 &     167 \\
\hline
\multicolumn{21}{c}{\bf Pulsars} \\
\hline
   1&  105759.0-522656 &10:57:59.00 &-52:26:56.57 & 0.04 &-0.69$\pm$0.01  &-0.62$\pm$0.02 &      $--$ &      $--$ &      $--$ &      $--$ &  15.22$\pm$ 0.10 & 105758-522655 &   21.0 $\pm$   1.2 &0.97 &      $------$ & $--$ & $--$ & $--$ &$--$ &     285 \\
\enddata
\end{deluxetable}

\begin{deluxetable}{llllllllllllllllllllll}
\tabletypesize{\tiny}
\rotate
\tablecaption{MGPS2/XMM: Extragalactic objects} 
\tablewidth{0pt}
\tablecolumns{2}
\tablehead{
\colhead{Src.} & 
\colhead{Name} &
\colhead{RA} &
\colhead{DEC} &
\colhead{err} &
\multicolumn{2}{c}{Hardness ratio} &
\multicolumn{4}{c}{Thermal -- Non thermal} &
\colhead{F$_{\rm x}$} &
\colhead{Name} &
\colhead{F$_{\rm r}$} &
\colhead{R$_{\rm xr}$} &
\colhead{Name} &
\multicolumn{3}{c}{2MASS Mag \& Colors} &
\colhead{R$_{\rm xi}$} &
\colhead{Redshift} &
\colhead{Bib.} \\ 
%}
\cline{8-11} \cline{17-19} 
%\tablehead{
\colhead{\#} & 
\colhead{2XMM J+} &
\colhead{h:m:s} &
\colhead{d:m:s} &
\colhead{''} &
\colhead{${{M-S}\over{S+M}}$} &
\colhead{${{H-M}\over{H+M}}$} &
\colhead{N$_{\rm H}$} &
\colhead{KT} &
\colhead{N$_{\rm H}$} &
\colhead{$\Gamma$} &
\colhead{[cgs]} &
\colhead{MGPS J+} &
\colhead{[mJy]} &
\colhead{f$_{<99}$\%} &
\colhead{2MASS J+} &
\colhead{$K$} &
\colhead{$H-K$} &
\colhead{$J-H$} &
\colhead{f$_{<99}$\%} &
\colhead{$z$} &
\colhead{Ref.}
}
\startdata
\hline
\multicolumn{22}{c}{\bf Normal galaxies} \\
\hline
   1&  154807.1-473818 &15:48: 7.15 &-47:38:18.43 & 0.22 & 0.71$\pm$0.02 &-0.03$\pm$0.02 &      0.8 &      2.6 &       1. &      2.2 &  94.45$\pm$ 4.00 & 154807-473821&   64.3 $\pm$   2.4 &1.23 & 154807-473818 &11.40 &0.742 &0.967 &1.162 &   $---$& 503-473 \\
   2&  161405.5-630837 &16:14: 5.59 &-63:08:37.13 & 1.67 & 0.33$\pm$0.03 &-0.40$\pm$0.03 &      0.4 &      1.6 &      0.6 &      3.0 &  23.15$\pm$ 1.48 & 161405-630827&   16.7 $\pm$   1.2 &2.88 & 161405-630837 &12.95 &0.379 &0.804 &2.988 &   $---$& 410-497 \\
\hline
\multicolumn{22}{c}{\bf Seyfert II} \\
\hline
   1&  163513.9-580447 &16:35:13.95 &-58:04:47.90 & 0.16 &-0.03$\pm$0.02 &-0.41$\pm$0.02 &      0.1 &      1.8 &      0.4 &      2.7 &   7.96$\pm$ 0.23 & 163513-580447&  111.9 $\pm$   3.6 &0.57 & 163513-580447 &11.42 &0.528 &0.860 &0.204 &   $---$&     392 \\
\hline
\multicolumn{22}{c}{\bf Cluster of galaxies} \\
\hline
   1&  132446.7-573630 &13:24:46.77 &-57:36:30.92 & 0.58 & 0.52$\pm$0.05 &-0.74$\pm$0.07 &       1. &     0.60 &     $--$ &     $--$ &   2.19$\pm$ 0.27 & 132446-573631& 3181.0 $\pm$ 126.6 &2.32 & 132446-573630 &11.61 &0.667 &1.133 &1.750 &   0.019&     299 \\
\enddata
\end{deluxetable}

\clearpage

\section{References}

\begin{itemize}

\item [1] Abazajian, K., et al.\ 2003, \aj, 126, 2081 
\item [2] Akiyama, M., Ueda, Y., Ohta, K., et al.\ 2003, \apjs, 148, 275 
\item [3] Akiyama, M.\ 2005, \apj, 629, 72 
\item [4] Aldcroft, T.~L., et al. \ 2003, \apj, 597, 751
\item [5] Alexander, D.~M., et al.\ 2003, \aj, 126, 539 
\item [6] Allen, S.~W., et al.\ 1992, \mnras, 259, 67 
\item [7] Allington-Smith, J.~R., et al. \ 1985, \mnras, 213, 243
\item [8] Anderson, M.~C., \& Rudnick, L.\ 1995, \apj, 441, 307 
\item [9] Andre, P., et al. \ 1987, \aj, 93, 1182 
\item [10] Awaki, H., et al.\ 2006, \apj, 645, 928 
\item [11] Badenes, C., et al.\ 2005, \apj, 624, 198 
\item [12] Baker, J.~R., et al. \ 1977, \aap, 59, 261 
\item [13] Baldi, A., et al. \ 2007, \apj, 666, 835 
\item [14] Balogh, M., et al. \ 2002, \mnras, 337, 256 
\item [15] Barber, T., et al. \ 2007, \mnras, 377, 787 
\item [16] Barcons, X., et al \ 2003, \mnras, 343, 137 
\item [17] Barcons, X., et al.\ 2007, \aap, 476, 1191 
\item [18] Barkhouse, W.~A., \& Hall, P.~B.\ 2001, \aj, 121, 2843 
\item [19] Barkhouse, W.~A., et al.\ 2006, \apj, 645, 955 
\item [20] Bastian, N., et al.\ 2005, \aap, 435, 65 
\item [21] Beasley, A.~J., et al.\ 2002, \apjs, 141, 13
\item [22] Beck, R.\ 2007, \aap, 470, 539 
\item [23] Becker, R.~H., et al.\ 1991, \apjs, 75, 1
\item [24] Becker, R.~H., et al.\ 1994, \apjs, 91, 347
\item [25] Becker, et al. \ 1995, \apj, 450, 559 
\item [26] Becker, W., et al. \ 1999, \aap, 352, 532
\item [27] Beckmann, V., et al. \ 2007, \aap, 475, 827 
\item [28] Belloni, T., et al.\ 1998, \aap, 339, 431 
\item [29] Belsole, E., et al.\ 2004, \mnras, 
\item [30] Belsole, E., et al. \ 2005, \mnras, 358, 120 
\item [31] Berkhuijsen, E.~M., et al.\ 1988, \aaps, 76, 65 
\item [32] Bernardi, M., et al.\ 2003, \aj, 125, 1817 
\item [33] Berrington, R.~C., et al. \ 2002, \aj, 123, 2261 
\item [34] Bettoni, D., et al. \ 2003, \aap, 399, 869 
\item [35] Bica, E., et al. \ 2003, \aap, 397, 177 
\item [36] Blomme, R., et al.\ 2007, \aap, 464, 701 
\item [37] B{\"o}hringer, H., et al.\ 2007, \aap, 469, 363
\item [38] Boller, T., et al. \ 1998, \aaps, 129, 87 
\item [39] Bonamente, M., et al. \ 2007, \apj, 668, 796 
\item [40] Bondi, M., et al.\ 2003, \aap, 403, 857
\item [41] Borkowski, K.~J., et al. \ 2007, \apjl, 671, L45 
\item [42] Bosch-Ramon, V., et al. \ 2007, \aap, 473, 545 
\item [43] Bournaud, F., et al. \ 2005, \aap, 438, 507 
\item [44] Boyle, B.~J., et al. \ 1997, \mnras, 285, 511 
\item [45] Branchesi, M., et al.\ 2007, \aap, 462, 449 
\item [46] Brassington, N.~J., et al.\ 2007, \mnras, 377, 1439 
\item [47] Bratsolis, E., et al.\ 2004, \aap, 423, 919 
\item [48] Braun, R., et al.\ 1990, \apjs, 72, 761 
\item [49] Brinkmann, W., et al. \ 1994, \aap, 281, 355 
\item [50] Brinkmann, W., et al.\ 1997, \aap, 323, 739 
\item [51] Brinkmann, W., et al. \ 2007, \aap, 463, 611
\item [52] Brittain, S.~D., et al. \ 2007, \apj, 659, 685 
\item [53] Brown, D.~L., \& Burns, J.~O.\ 1991, \aj, 102, 1917
\item [54] Brown, M.~J.~I., et al. \ 2001, \aj, 121, 2381
\item [55] Brunthaler, A., et al. \ 2007, \aap, 462, 101 
\item [56] Buote, D.~A., \& Canizares, C.~R.\ 1998, \mnras, 298, 811 
\item [57] Burenin, R.~A., et al. \ 2007, \apjs, 172, 561
\item [58] Buta, R.\ 1995, \apjs, 96, 39 
\item [59] Caccianiga, A., et al.\ 2000, \aaps, 144, 247 
\item [60] Caccianiga, A., et al. \ 2007, \aap, 470, 557 
\item [61] Camilo, F., et al.\ 2006, \apj, 637, 456 
\item [62] Cardamone, C.~N., et al. \ 2007, \aj, 134, 1263 
\item [63] Carilli, C.~L., et al.\ 2004, \aj, 128, 997 
\item [64] Carpenter, J.~M., et al.\ 2002, \aj, 124, 1001 
\item [65] Carraro, G.\ 2002, \mnras, 331, 785
\item [66] Carrasco, E.~R., et al.\ 2007, \apj, 664, 777 
\item [67] Carrera, F.~J., et al.\ 2007, \aap, 469, 27 
\item [68] Caswell, J.~L.\ 1998, \mnras, 297, 215 
\item [69] Chambers, K.~C., et al.\ 1996, \apjs, 106, 247
\item [70] Chand, H., et al.\ 2004, \aap, 417, 853
\item [71] Chand, H., et al.\ 2005, \aap, 430, 47 
\item [72] Chatterjee, S., et al.\ 2005, \apjl, 634, L101 
\item [73] Chatzikos, M., et al. \ 2006, \apj, 643, 751 
\item [74] Chen, C.-H.~R., et al.\ 2005, \apj, 619, 779 
\item [75] Chen, P.-S., et al., Chinese Journal of A\&A, 7, 657 
\item [76] Chiappetti, L., et al.\ 2005, \aap, 439, 413 
\item [77] Christlein, D., \& Zabludoff, A.~I.\ 2003, \apj, 591, 764 
\item [78] Ciliegi, P., et al. \ 1997, \mnras, 284, 401 
\item [79] Cocchia, F., et al.\ 2007, \aap, 466, 31
\item [80] Condon, J.~J., et al. \ 1990, \aj, 99, 1071 
\item [81] Condon, J.~J., et al. \ 1998, \aj, 115, 1693
\item [82] Conselice, C.~J., \& Gallagher, J.~S., III 1999, \aj, 117, 75 
\item [83] Corwin, H.~G.\ 1995, NED Team Report, 1, 1 (1995), 1, 1 
\item [84] Cotton, W.~D., et al. \ 1999, \apjs, 125, 409
\item [85] Couch, W.~J., et al. \ 1991, \mnras, 249, 606 
\item [86] Crawford, C.~S.\ 2006, Astronomische Nachrichten, 327, 191
\item [87] Croom, S.~M., et al. \ 2001, \mnras, 322, L29 
\item [88] Cunow, B., \& Wargau, W.~F.\ 1993, \aaps, 102, 331 
\item [89] Cutri, R.~M., et al.\ 2003, The IRSA 2MASS ASPS Catalog, NASA/IPAC  
\item [90] Dadina, M.\ 2007, \aap, 461, 1209 
\item [91] Dahlem, M., \& Thiering, I.\ 2000, \pasp, 112, 148 
\item [92] Damiani, F., et al. \ 2004, \apj, 608, 781 
\item [93] Danner, R.\ 1998, \aaps, 128, 331 
\item [94] Darnley, M.~J., et al.\ 2007, \apjl, 661, L45 
\item [95] Davis, D.~S., et al. \ 2003, \apj, 597, 202 
\item [96] De Breuck, C., et al. \ 2002, \aap, 394, 59 
\item [97] de Grijs, R., et al. \ 2005, \mnras, 361, 311 
\item [98] de la Varga, A., et al. \ 2000, \aap, 363, 69 
\item [99] De Propris, R., et al.\ 2002, \mnras, 329, 87 
\item [100] de Ruiter, H.~R., et al.\ 1998, \aap, 339, 34 
\item [101] de Vries, W.~H., et al. \ 2004, \aj, 127, 2565
\item [102] Dewdney, P.~E., et al.\ 1991, \apjs, 76, 1055 
\item [103] Dickey, J.~M., \& Salpeter, E.~E.\ 1984, \apj, 284, 461 
\item [104] Dickey, J.~M., et al.\ 2000, \apj, 536, 756 
\item [105] Dodson, R.~G., \& Ellingsen, S.~P.\ 2002, \mnras, 333, 307
\item [106] Donnelly, R.~H., et al. \ 2001, \apj, 562, 254 
\item [107] Douglas, J.~N., et al. \ 1996, \aj, 111, 1945 
\item [108] Downes, D., et al. \ 1993, \apjl, 414, L13 
\item [109] Downes, R., et al. \ 1997, \pasp, 109, 345 
\item [110] Drake, S.~A.\ 1990, \aj, 100, 572 
\item [111] Dudik, R.~P., et al.\ 2005, \apj, 620, 113 
\item [112] Duric, N., et al. \ 1995, \apj, 445, 173
\item [113] Durret, F., et al. \ 2005, \aap, 432, 809 
\item [114] Dwelly, T., \& Page, M.~J.\ 2006, \mnras, 372, 1755 
\item [115] Eckart, M.~E., et al. \ 2005, \apjs, 156, 35
\item [116] Ellis, S.~C., \& O'Sullivan, E.\ 2006, \mnras, 367, 627 
\item [117] Esposito, P., et al.\ 2007, \aap, 467, L45 
\item [118] Evans, I.~N., et al. \ 1996, \apjs, 105, 93 
\item [119] Evans, A.~S., et al. \ 2002, \apj, 580, 749 
\item [120] Fadda, D., et al. \ 2002, \aap, 383, 838
\item [121] Falco, E.~E., et al.\ 1999, \pasp, 111, 438
\item [122] Falgarone, E., \& Gilmore, W.\ 1981, \aap, 95, 32
\item [123] Falomo, R.\ 1991, \aj, 101, 821 
\item [124] Fanti, C., et al.\ 2001, \aap, 369, 380
\item [125] Feretti, L., et al.\ 2006, \mnras, 368, 544 
\item [126] Ferrari, C., et al. \ 2006, \aap, 457, 21 
\item [127] Ficarra, A., et al. \ 1985, \aaps, 59, 255
\item [128] Filho, M.~E., et al. \ 2002, \apjs, 142, 223 
\item [129] Filipovi{\'c}, M. D., et al. \ 2000, \aap, 353,
\item [130] Filipovi{\'c}, M.~D., et al.\ 2002, \mnras, 335, 1085 
\item [131] Finoguenov, A., et al.\ 2006, \apj, 646, 143
\item [132] Finoguenov, A., et al.\ 2007, \apjs, 172, 182 
\item [133] Flohic, H.~M.~L.~G., et al. \ 2006, \apj, 647, 140
\item [134] Fomalont, E.~B., et al. \ 2003, \aj, 126, 2562
\item [135] Foschini, L., et al.\ 2002, \aap, 392, 817 
\item [136] Foschini, L., et al.\ 2006, \aap, 453, 829
\item [137] Fujita, Y., et al. \ 2006, \pasj, 58, 131 
\item [138] Gaetz, T.~J., et al.\ 2007, \apj, 663, 234 
\item [139] Galbiati, E., et al.\ 2005, \aap, 430, 927 
\item [140] Gallo, L.~C.\ 2006, \mnras, 365, 960
\item [141] Gallo, L.~C., et al. \ 2006, \mnras, 365, 688 
\item [142] Gandhi, P., et al. \ 2004, \mnras, 348, 529 
\item [143] Garnier, R., et al.\ 1996, \aaps, 117, 467 
\item [144] Gastaldello, F., et al. \ 2007, \apj, 669, 158
\item [145] Gavazzi, G.\ 1979, \aap, 72, 1 
\item [146] Gavazzi, G., \& Contursi, A.\ 1994, \aj, 108, 24 
\item [147] Geller, M.~J., et al. \ 2006, \aj, 132, 2243 
\item [148] Genzel, R., et al. \ 2001, \apj, 563, 527 
\item [149] Georgakakis, A.~E., et al.\ 2004, \mnras, 354, 123
\item [150] Georgakakis, A., et al.\ 2004, \mnras, 349, 135 
\item [151] Georgakakis, A.~E., et al.\ 2006, \mnras, 367, 1017 
\item [152] Georgantopoulos, I., et al.\ 2004, \apj, 614, 634 
\item [153] Georgantopoulos, I., \& Georgakakis, A. 2005, \mnras, 358, 131 
\item [154] Georgantopoulos, I., et al.\ 2005, \mnras, 360, 782
\item [155] Getman, K.~V., et al.\ 2005, \apjs, 160, 319
\item [156] Getman, K.~V., et al. \ 2007, \apj, 654, 316 
\item [157] Gil, J., et al. \ 2006, \apj, 650, 1048
\item [158] Gilbank, D.~G., et al. \ 2004, \mnras, 348, 551 
\item [159] Girart, J.~M., et al. \ 2002, RMA\&A, 38, 169 
\item [160] Godwin, J.~G., et al.\ 1983, \mnras, 202, 113 
\item [161] Gondoin, P.\ 2005, \aap, 438, 291 
\item [162] Gondoin, P.\ 2006, \aap, 454, 595
\item [163] Gordon, S.~M., et al. \ 1999, \apjs, 120, 247 
\item [164] Goss, W.~M., et al.\ 1982, \mnras, 198, 259 
\item [165] Gosset, E., et al. \ 2005, \aap, 429, 685
\item [166] Grainge, K., et al. \ 1996, \mnras, 278, L17 
\item [167] Gratton, R.~G., et al.\ 2007, \aap, 464, 953 
\item [168] Gregory, P.~C., et al. \ 1996, \apjs, 103, 427 
\item [169] Griffith, M.~R., et al. \ 1994, \apjs, 90, 179 
\item [170] Griffith, M.~R., et al. \ 1995, \apjs, 97, 347 
\item [171] Grimes, J.~P., et al.\ 2007, \apj, 668, 891 
\item [172] Gruendl, R.~A., et al. \ 2006, \apj, 653, 339 
\item [173] Grupe, D., et al.\ 2004, \aj, 127, 156
\item [174] Grupe, D., et al. \ 2006, \aj, 131, 55 
\item [175] Gruppioni, C., et al.\ 1997, \mnras, 286, 470 
\item [176] Gruppioni, C., et al. \ 1999, \mnras, 304, 199 
\item [177] Gruppioni, C., et al.\ 1999, \mnras, 305, 297
\item [178] Guainazzi, M., et al.\ 2005, \aap, 444, 119 
\item [179] Guainazzi, M., et al. \ 2006, \aap, 446, 87 
\item [180] Guainazzi, M., \& Bianchi, S.\ 2007, \mnras, 374, 1290 
\item [181] G{\"u}del, M., et al.\ 2007, \aap, 468, 353
\item [182] Guerrero, M.~A., et al.\ 2004 BAAS, 36, 1570 
\item [183] Haberl, F., et al. \ 2000, \aaps, 142, 41 
\item [184] Harris, D.~E., et al. \ 1980, \aaps, 39, 215 
\item [185] Helfand, D.~J., \& Chanan, G.~A.\ 1989, \aj, 98, 1652
\item [186] Helou, G., \& Walker, D.W.\ 1988, IAS (IRAS) c. V 7, 1-265 
\item [187] Hands, A.~D.~P., et al. \ 2004, \mnras, 351, 31 
\item [188] Harrison, F.~A., et al. \ 2003, \apj, 596, 944 
\item [189] Hatziminaoglou, E., et al.\ 2002, \aap, 384, 81
\item [190] Hayakawa, A., et al.\ 2006, \pasj, 58, 695 
\item [191] Helsdon, S.~F., et al. \ 2001, \mnras, 325, 693 
\item [192] Hirabayashi, H., et al.\ 2000, \pasj, 52, 997 
\item [193] Ho, L.~C., \& Ulvestad, J.~S.\ 2001, \apjs, 133, 77 
\item [194] Hoff, W., et al. \ 1998, \aap, 336, 242 
\item [195] Hook, I.~M., et al. \ 2002, \aap, 391, 509 
\item [196] Hornschemeier, et al.\ 2005, \aj, 129, 86 
\item [197] Hornschemeier, A.~E., et al.\ 2006, \apj, 643, 144 
\item [198] Hota, A., \& Saikia, D.~J.\ 2005, \mnras, 356, 998 
\item [199] Huang, H.~H., \& Becker, W.\ 2007, \aap, 463, L5 
\item [200] Huchra, J., et al.\ 1993, \aj, 105, 1637 
\item [201] Hudaverdi, M., et al.\ 2006, \pasj, 58, 931 
\item [202] Hui, C.~Y., \& Becker, W.\ 2007, \aap, 467, 1209 
\item [203] Humphrey, P.~J., \& Buote, D.~A.\ 2004, \apj, 612, 848 
\item [204] Hwang, H.~S., et al.\ 2007, \mnras, 375, 115
\item [205] Hyman, S.~D., et al.\ 2001, \apj, 551, 702 
\item [206] Immler, S., \& Kuntz, K.~D.\ 2005, \apjl, 632, L99 
\item [207] Indebetouw, R., et al.\ 2004, \aj, 128, 2206 
\item [208] Israel, F.~P., \& van der Kruit, P.~C.\ 1974, \aap, 32, 363
\item [209] Israel, F.~P., \& Koornneef, J.\ 1991, \aap, 248, 404 
\item [210] Ivanov, G.~R., et al.\ 1993, \apjs, 89, 85 
\item [211] Jackson, C.~A., et al.\ 2002, \aap, 386, 97 
\item [212] Jaffe, W.~J., \& Perola, G.~C.\ 1975, \aaps, 21, 137 
\item [213] James, P.~A., \& Anderson, J.~P.\ 2006, \aap, 453, 57 
\item [214] Jeffries, R.~D., et al.\ 2006, \mnras, 367, 781 
\item [215] Jenkins, L.~P., et al.\ 2005, \mnras, 357, 109 
\item [216] Jim{\'e}nez-Bail{\'o}n, E., et al.\ 2007, \aap, 469, 881 
\item [217] Joncas, G., et al.\ 1985, \apj, 298, 596
\item [218] Jones, P.~A., \& McAdam, W.~B.\ 1992, \apjs, 80, 137 
\item [219] Kaastra, J.~S., et al.\ 2004, \aap, 413, 415 
\item [220] Kadler, M., et al.\ 2005, Astronomische Nachrichten, 326, 545 
\item [221] Kahabka, P., et al.\ 1999, \aaps, 136, 81 
\item [222] Kaldare, R., et al.\ 2003, \mnras, 339, 652 
\item [223] Kassim, N.~E., et al.\ 2007, \apjs, 172, 686 
\item [224] Katgert, P., et al.\ 1998, \aaps, 129, 399 
\item [225] Kennicutt, R.~C., et al.\ 2003, \pasp, 115, 928 
\item [226] Kerber, F., et al.\ 2003, \aap, 408, 1029 
\item [227] Kharb, P., et al.\ 2008, \apjs, 174, 74
\item [228] Kim, K.-T., et al.\ 1994, \aaps, 105, 385 
\item [229] Kim, K.-T., \& Koo, B.-C.\ 2001, \apj, 549, 979
\item [230] Kim, D.-W., et al.\ 2004, \apjs, 150, 19
\item [231] Kollgaard, R.~I., et al.\ 1994, \apjs, 93, 145 
\item [232] Kolokotronis, V., et al.\ 2006, \mnras, 366, 163 
\item [233] Kondratko, P.~T., et al.\ 2005, \apj, 618, 618 
\item [234] Kong, A.~K.~H.\ 2003, \mnras, 346, 265 
\item [235] Koulouridis, E., et al.\ 2006, \apj, 651, 93 
\item [236] Krautter, J., et al.\ 1997, \aaps, 123, 329 
\item [237] Krawczynski, H., et al.\ 2003, \mnras, 345, 1255 
\item [238] Lamm, M.~H., et al.\ 2004, \aap, 417, 557 
\item [239] Landecker, T.~L., et al.\ 1989, \mnras, 237, 277
\item [240] Landt, H., et al.\ 2001, \mnras, 323, 757 
\item [241] Lane, K.~P., et al.\ 2007, \mnras, 378, 716 
\item [242] La Palombara, N., et al.\ 2006, \aap, 458, 245 
\item [243] Large, M.~I., et al.\ 1981, \mnras, 194, 693 
\item [244] Large, M.~I., et al.\ 1991, The Observatory, 111, 72
\item [245] LaRosa, T.~N., et al.\ 2000, \aj, 119, 207 
\item [246] Larsen, S.~S.\ 1999, \aaps, 139, 393
\item [247] Larsen, S.~S.\ 2004, \aap, 416, 537 
\item [248] Lauberts, A.\ 1982, Garching: (ESO), 1982,  
\item [249] Laurent-Muehleisen, S.~A., et al.\ 1997, \aaps, 122, 235 
\item [250] Lawrence, C.~R., et al.\ 1986, \apjs, 61, 105
\item [251] Lawrence, A., et al.\ 1999, \mnras, 308, 897
\item [252] Lazio, T.~J.~W., \& Cordes, J.~M.\ 1998, \apjs, 118, 201 
\item [253] Leahy, J.~P., \& Perley, R.~A.\ 1995, \mnras, 277, 1097
\item [254] Le Campion, J.-F., et al.\ 1992, \aaps, 95, 233 
\item [255] Lehtinen, K., et al.\ 2003, \aap, 401, 1017 
\item [256] Lira, P., et al.\ 2002, \mnras, 330, 259 
\item [257] Liske, J., et al.\ 2003, \mnras, 344, 307 
\item [258] Liu, J.-F., \& Bregman, J.~N.\ 2005, \apjs, 157, 59 
\item [259] Liu, Q.~Z., et al.\ 2007, \aap, 469, 807 
\item [260] Longinotti, A.~L., et al.\ 2007, \aap, 470, 73
\item [261] Lorimer, D.~R., et al.\ 2006, \mnras, 372, 777
\item [262] Loveday, J. \ 1996, \mnras, 278, 1025
\item [263] Loveday, J., et al.\ 1996, \apjs, 107, 201 
\item [264] Lu, N.~Y., \& Freudling, W.\ 1995, \apj, 449, 527 
\item [265] Ma, C., et al.\ 1998, \aj, 116, 516 
\item [266] Machalski, J., \& Condon, J.~J.\ 1999, \apjs, 123, 41 
\item [267] Mack, K.-H., et al.\ 1997, \aaps, 123, 423 
\item [268] Magazzu, A., et al.\ 1997, \aaps, 124, 449 
\item [269] Magliocchetti, M., et al.\ 2002, \mnras, 333, 100
\item [270] Magliocchetti, M., et al.\ 2004, \mnras, 350, 1485
\item [271] Manchester, R.~N., et al.\ 2001, \mnras, 328, 17 
\item [272] Marx, M., et al.\ 1997, \aaps, 126, 325
\item [273] Marx-Zimmer, M., et al.\ 2000, \aap, 354, 787 
\item [274] Mart\'{\i}i, J., et al.\ 1993, \apj, 416, 208 
\item [275] Marvel, K.~B., et al.\ 1999, \apjs, 120, 147
\item [276] Maslowski, J., \& Kellermann, K.~I.\ 1988, \aj, 95, 1659 
\item [277] Mason, K.~O., et al.\ 2000, \mnras, 311, 456 
\item [278] Massey, P., \& Thompson, A.~B.\ 1991, \aj, 101, 1408
\item [279] Massey, P., \& Johnson, O.\ 1998, \apj, 505, 793 
\item [280] Mateos, S., et al.\ 2005, \aap, 444, 79
\item [281] Mauch, T., et al.\ 2003, \mnras, 342, 1117
\item [282] Maughan, B.~J., et al.\ 2004, \mnras, 354, 1 
\item [283] McCarthy, P.~J.\ 1991, \aj, 102, 518
\item [284] McCarthy, P.~J., et al.\ 1996, \apjs, 107, 19
\item [285] McGowan, K.~E., et al.\ 2006, \apj, 639, 377 
\item [286] McKay, N.~P.~F., et al.\ 2004, \mnras, 352, 1121
\item [287] Meyssonnier, N., et al.\ 1993, \aaps, 102, 251 
\item [288] Miller, N.~A.\ 2005, \aj, 130, 2541
\item [289] Mingaliev, M.~G., et al.\ 2001, \aap, 370, 78
\item [290] Miyauchi-Isobe, N., \& Maehara, H.\ 1998, PNAO of Japan, 5, 75
\item [291] Moretti, A., et al.\ 2004, \aap, 428, 21
\item [292] Moshir, M., et al.\ 1990, IRAS FS Catalogue, V2.0 (1990), 0 
\item [293] Mould, J., et al.\ 2004, \apjs, 154, 623 
\item [294] Moy, E., et al.\ 2003, \aap, 403, 493 
\item [295] Muchovej, S., et al.\ 2007, \apj, 663, 708 
\item [296] Mullis, C.~R., et al.\ 2003, \apj, 594, 154 
\item [297] Muno, M.~P., et al.\ 2004, \apj, 613, 1179 
\item [298] Mu{\~n}oz, J.~A., et al.\ 2003, \apj, 594, 684
\item [299] Nagayama, T., et al.\ 2006, \mnras, 368, 534 
\item [300] Nakazawa, K., et al.\ 2007, \pasj, 59, 167
\item [301] Naz{\'e}, Y., et al.\ 2004, \apj, 608, 208 
\item [302] Naz{\'e}, Y., et al.\ 2004, \aap, 417, 667 
\item [303] Naz{\'e}, Y., et al.\ 2004, \aap, 418, 841 
\item [304] NED Team 1992, NED Team Report, 1, 1 (1992), 1, 1 
\item [305] Nolan, L.~A., et al.\ 2004, \mnras, 353, 221 
\item [306] Nordon, R., \& Behar, E.\ 2007, \aap, 464, 309 
\item [307] Novara, G., et al.\ 2006, \aap, 448, 93 
\item [308] Odewahn, S.~C., \& Aldering, G.\ 1995, \aj, 110, 2009 
\item [309] Oegerle, W.~R., et al.\ 1991, \apj, 376, 46 
\item [310] Oey, M.~S.\ 1996, \apjs, 104, 71 
\item [311] Ohyama, Y., et al.\ 1999, \aj, 117, 2617 
\item [312] O'Meara, J.~M., et al.\ 2007, \apj, 656, 666
\item [313] Ott, J., et al.\ 2005, \mnras, 358, 1423 
\item [314] Ott, J., et al.\ 2005, \mnras, 358, 1423 
\item [315] Owen, F.~N., et al.\ 1993, \apjs, 87, 135
\item [316] Owen, F.~N., et al.\ 1995, \aj, 109, 14 
\item [317] Page, M.~J., et al.\ 2001, \mnras, 325, 575 
\item [318] Page, K.~L., et al.\ 2005, \mnras, 364, 195 
\item [319] Page, M.~J., et al.\ 2006, \mnras, 369, 156 
\item [320] Page, M.~J., et al.\ 2007, \mnras, 378, 1335 
\item [321] Panessa, F., et al.\ 2006, \aap, 455, 173
\item [322] Panessa, F., et al.\ 2007, \aap, 467, 519
\item [323] Papadakis, I.~E., et al.\ 2007, \aap, 461, 931 
\item [324] Parker, Q.~A., et al.\ 2006, \mnras, 373, 79 
\item [325] Pasquini, L., \& Belloni, T.\ 1998, \aap, 336, 902 
\item [326] Paturel, G., et al.\ 1989, \aaps, 80, 299 
\item [327] Paturel, G., et al.\ 1999, \aaps, 140, 89 
\item [328] Paturel, G., et al.\ 2000, \aaps, 144, 475 
\item [329] Paturel, G., et al.\ 2002, LEDA, 0 (2002), 0 
\item [330] Pavlov, G.~G., et al.\ 2007, \apj, 664, 1072 
\item [331] Payne, J.~L., et al.\ 2004, \aap, 425, 443 
\item [332] Perlman, E.~S., et al.\ 2002, \apjs, 140, 265 
\item [333] Perola, G.~C., et al.\ 2004, \aap, 421, 491
\item [334] Perryman, M.~A.~C., et al.\ 1997, \aap, 323, L49
\item [335] Pestalozzi, M.~R., et al.\ 2005, \aap, 432, 737 
\item [336] Piconcelli, E., et al.\ 2002, \aap, 394, 835
\item [337] Piconcelli, E., et al.\ 2003, \aap, 412, 689 
\item [338] Piconcelli, E., et al.\ 2005, \aap, 432, 15
\item [339] Piconcelli, E., et al.\ 2007, \aap, 466, 855 
\item [340] Piatek, S., et al.\ 2005, \aj, 130, 95 
\item [341] Pietsch, W., \& Arp, H.\ 2001, \aap, 376, 393 
\item [342] Pietsch, W., et al.\ 2004, \aap, 426, 11
\item [343] Pietsch, W., et al.\ 2005, \aap, 434, 483
\item [344] Pillitteri, I., et al.\ 2006, \aap, 450, 993 
\item [345] Plionis, M., et al.\ 2005, \apjl, 622, L17 
\item [346] Pollanen, M.~D., \& Feldman, P.~A.\ 1995, \pasp, 107, 617 
\item [347] Polletta, M.~d.~C., et al.\ 2006, \apj, 642, 673
\item [348] Polletta, M., et al.\ 2007, \apj, 663, 81 
\item [349] Pollock, A.~M.~T.\ 2007, \aap, 463, 1111
\item [350] Pont, F., et al.\ 2004, \aj, 127, 840 
\item [351] Pooley, D., et al.\ 2002, \apj, 573, 184 
\item [352] Pooley, D., et al.\ 2007, \apj, 661, 19
\item [353] Postman, M., et al. \ 1996, \aj, 111, 615 
\item [354] Prandoni, I., et al.\ 2007, New Astronomy Review, 51, 43
\item [355] Proust, D., et al.\ 2000, \aap, 355, 443 
\item [356] Quintana, H., \& Ramirez, A.\ 1995, \apjs, 96, 343 
\item [357] Raassen, A.~J.~J., et al.\ 2007, \mnras, 379, 1075 
\item [358] Radburn-Smith, D.~J., et al.\ 2006, \mnras, 369, 1131 
\item [359] Raiteri, C.~M., et al.\ 2007, \aap, 473, 819
\item [360] Rakowski, C.~E., et al.\ 2006, \apjl, 649, L111 
\item [361] Rajagopal, J., et al.\ 1998, JA\&A, 19, 97
\item [362] Ramella, M., et al.\ 2002, \aj, 123, 2976 
\item [363] Ram{\'{\i}}rez, S.~V., et al.\ 2004, \aj, 127, 2659
\item [364] Randich, S., et al.\ 1995, \aap, 300, 134
\item [365] Ratcliffe, A., et al.\ 1998, \mnras, 300, 417 
\item [366] Read, A.~M., et al.\ 1997, \mnras, 286, 626 
\item [367] Read, A.~M., \& Pietsch, W.\ 2001, \aap, 373, 473
\item [368] Read, A.~M.\ 2005, \mnras, 359, 455
\item [369] Rebull, L.~M.\ 2001, \aj, 121, 1676 
\item [370] Reid, A.~D., et al.\ 1998, \mnras, 296, 531 
\item [371] Rengelink, R.~B., et al.\ 1997, \aaps, 124, 259 
\item [372] Reynoso, E.~M., et al.\ 2006, \aap, 449, 243 
\item [373] Rho, J., et al.\ 2007, \apj, 666, 1108
\item [374] Risaliti, G., et al.\ 2007, \apjl, 659, L111
\item [375] Roettgering, H.~J.~A., et al.\ 1994, \aaps, 108, 79 
\item [376] Romer, A.~K., et al.\ 2000, \apjs, 126, 209 
\item [377] Rossetti, M., et al.\ 2007, \aap, 463, 839 
\item [378] Ruiz, A., et al.\ 1992, \apj, 398, 139 
\item [379] Ruiz, A., et al.\ 2007, \aap, 471, 775
\item [380] Rupke, D.~S., et al.\ 2005, \apjs, 160, 87
\item [381] Ryabinkov, A.~I., et al.\ 2003, \aap, 412, 707 
\item [382] Ryder, S., et al.\ 1993, \apj, 416, 167 
\item [383] Salpeter, E.~E., \& Dickey, J.~M.\ 1987, \apj, 317, 102 
\item [384] Saikia, D.~J., et al.\ 2007, \mnras, 375, L31
\item [385] Saintonge, A., et al.\ 2005, \apjs, 157, 228 
\item [386] Sambruna, R.~M., et al. \ 2007, \apj, 669, 884 
\item [387] Sana, H., et al.\ 2006, \aap, 454, 1047 
\item [388] S{\'a}nchez-Sutil, J.~R., et al.\ 2006, \aap, 452, 739 
\item [389] Sansom, A.~E., et al.\ 2006, \mnras, 370, 1541 
\item [390] Sasaki, M., et al.\ 2000, \aaps, 143, 391 
\item [391] Sato, K., et al.\ 2005, \pasj, 57, 743 
\item [392] Sazonov, S., et al.\ 2007, \aap, 462, 57 
\item [393] Schartel, N., et al.\ 2007, \aap, 474, 431
\item [394] Schild, H., et al.\ 2003, \aap, 397, 859 
\item [395] Schindler, S.\ 2000, \aaps, 142, 433
\item [396] Schlegel, E.~M., et al.\ 2000, \aj, 120, 2373 
\item [397] Schmidt, K.-H., \& Boller, T.\ 1992, Astro. Nach. 313, 189 
\item [398] Schmitt, H.~R., et al.\ 2001, \apjs, 132, 199
\item [399] Sciortino, V.~L., \& Martin, C.~L.\ 2004, BAAS, 36, 1387 
\item [400] Scott, J.~E., et al.\ 2005, \apj, 634, 193
\item [401] Sengupta, C., \& Balasubramanyam, R.\ 2006, \mnras, 369, 360 
\item [402] Severgnini, P., et al.\ 2005, \aap, 431, 87 
\item [403] Seymour, N., et al.\ 2007, \apjs, 171, 353 
\item [404] Shectman, S.~A., et al.\ 1996, \apj, 470, 172 
\item [405] Shtykovskiy, P., \& Gilfanov, M.\ 2005, \aap, 431, 597 
\item [406] Sidoli, L., et al.\ 2006, \aap, 459, 901
\item [407] Silverman, J.~D., et al.\ 2005, \apj, 618, 123
\item [408] Singh, K.~P.\ 1999, \mnras, 309, 991 
\item [409] Skinner, S.~L., et al.\ 2007, \apj, 658, 1144 
\item [410] Skrutskie, M.~F., et al.\ 2006, \aj, 131, 1163 
\item [411] Slee, O.~B., et al.\ 1982, PASA, 4, 278 
\item [412] Slee, O.~B., et al.\ 1994, AJP, 47, 145 
\item [413] Slee, O.~B., et al.\ 1996, AJP, 49, 977 
\item [414] Slee, O.~B., et al.\ 2004, PASA, 21, 72 
\item [415] Slezak, E., et al.\ 1998, \aaps, 128, 67 
\item [416] Smith, E.~O., et al.\ 1997, \aj, 114, 1471 
\item [417] Smith, D.~A., et al.\ 1998, \apj, 494, 150 
\item [418] Smith, R.~J., et al.\ 2004, \aj, 128, 1558 
\item [419] Soderberg, A.~M., et al.\ 2007, \apj, 661, 982
\item [420] Sokoloski, J.~L., et al.\ 1996, \apj, 459, 142 
\item [421] Soria, R., \& Wong, D.~S.\ 2006, \mnras, 372, 1531 
\item [422] Spangler, S.~R., \& Cordes, J.~M.\ 1998, \apj, 505, 766 
\item [423] Sramek, R.\ 1992, RB. AGN \& Starburst Galaxies, 31, 273 
\item [424] Stauffer, J.~R., et al.\ 1994, \apjs, 91, 625
\item [425] Steffen, A.~T., et al.\ 2006, \aj, 131, 2826
\item [426] Stein, P.\ 1996, \aaps, 116, 203 
\item [427] Stern, D., et al.\ 2002, \aj, 123, 2223
\item [428] Stickel, M., et al.\ 2004, \aap, 422, 39 
\item [429] Strateva, I.~V., et al.\ 2005, \aj, 130, 387
\item [430] Strateva, I.~V., et al. \ 2006, \apj, 651, 749
\item [431] Strickland, D.~K.\ 2007, \mnras, 376, 523 
\item [432] Strom, R.~G., et al.\ 1990, \aap, 227, 19 
\item [433] Suh, J.~A., et al.\ 2005, \apj, 630, 1074 
\item [434] Supper, R., et al.\ 1997, \aap, 317, 328 
\item [435] Sun, M., et al.\ 2007, \apj, 657, 197 
\item [436] Sutaria, F.~K., et al.\ 2003, \aap, 397, 1011 
\item [437] Swartz, D.~A., et al.\ 2004, \apjs, 154, 519 
\item [438] Szymczak, M., et al.\ 2000, \aaps, 143, 269 
\item [439] Tajer, M., et al.\ 2005, \aap, 435, 799 
\item [440] Tajer, M., et al.\ 2007, \aap, 467, 73 
\item [441] Taylor, A.~R., et al.\ 1996, \apjs, 107, 239 
\item [442] Taylor, G.~B., et al.\ 2005, \apjs, 159, 27 
\item [443] Telleschi, A., et al.\ 2007, \aap, 468, 443 
\item [444] Teng, S.~H., et al. \ 2005, \apj, 633, 664
\item [445] Terashima, Y., \& Wilson, A.~S.\ 2004, \apj, 601, 735 
\item [446] Terlevich, R., et al. \ 1991, \aaps, 91, 285 
\item [447] Thomas, T., \& Katgert, P.\ 2006, \aap, 446, 19 
\item [448] Thompson, D., et al.\ 1994, \aj, 108, 828
\item [449] Tinney, C.~G., et al.\ 1997, \mnras, 285, 111 
\item [450] Torres, D.~F., et al.\ 2006, \aap, 457, 257 
\item [451] Tramacere, A., et al.\ 2007, \aap, 467, 501 
\item [452] Treister, E., et al.\ 2005, \apj, 621, 104
\item [453] Treu, T., et al. \ 2003, \apj, 591, 53 
\item [454] Treves, A., et al.\ 2007, \aap, 473, L17 
\item [455] Trushkin, S.~A., et al.\ 1987, AIISAO, 25, 84 
\item [456] Tsvetanov, Z.~I., \& Petrosian, A.~R.\ 1995, \apjs, 101, 287 
\item [457] Turner, J.~L., \& Ho, P.~T.~P.\ 1994, \apj, 421, 122 
\item [458] Tytler, D., et al.\ 2004, \aj, 128, 1058
\item [459] Ueda, Y., et al.\ 1999, \apj, 518, 656 
\item [460] Uro{\v s}evi{\'c}, D., et al.\ 2005, \aap, 435, 437
\item [461] Vader, J.~P., et al.\ 1993, \aj, 106, 1743 
\item [462] Vald{\'e}s, J.~R., et al. \ 2005, \aap, 434, 149 
\item [463] van den Bergh, S., et al.\ 2003, \pasp, 115, 1280
\item [464] van der Hulst, J.~M.\ 1979, \aap, 75, 97 
\item [465] Van Dyk, S.~D., et al.\ 2003, \pasp, 115, 448 
\item [466] van Vliet, W., et al.\ 1976, \aap, 47, 345 
\item [467] Venturi, T., et al.\ 2000, \mnras, 314, 594
\item [468] Venturi, T., et al.\ 2002, \aap, 385, 39
\item [469] Verbunt, F.\ 2001, \aap, 368, 137 
\item [470] Vig, S., et al.\ 2006, \aap, 446, 1021 
\item [471] Vink, J., et al.\ 2006, \mnras, 367, 928
\item [472] Visser, A.~E., et al.\ 1995, \aaps, 110, 419
\item [473] Visvanathan, N., \& Yamada, T.\ 1996, \apjs, 107, 521
\item [474] Voges, W., et al.\ 1999, \aap, 349, 389
\item [475] Vogler, A., et al.\ 1996, \aap, 305, 74 
\item [476] Vogler, A., \& Pietsch, W.\ 1999, \aap, 352, 64 
\item [477] V{\'e}ron-Cetty, M.-P., et al.\ 2004, \aap, 414, 487 
\item [478] V{\'e}ron-Cetty, M.-P., \& V{\'e}ron, P.\ 2006, \aap, 455, 773 
\item [479] Walter, F., et al.\ 2002, \aj, 123, 225 
\item [480] Walterbos, R.~A.~M.,.\ 1985, \aaps, 61, 451 
\item [481] Wambsganss, J., et al.\ 1999, \aap, 346, L5
\item [482] Wang, J.~X., et al.\ 2007, \apj, 657, 95 
\item [483] Waskett, T.~J., et al.\ 2004, \mnras, 350, 785 
\item [484] Watson, D., et al.\ 1999, \aap, 345, 414 
\item [485] Watson, M.~G., et al.\ 2005, \aap, 437, 899 
\item [486] Wegner, G., et al.\ 2001, \aj, 122, 2893
\item [487] Weisskopf, M.~C., et al.\ 2006, \apj, 652, 387 
\item [488] White, R.~L., et al.\ 1997, \apj, 475, 479
\item [489] White, R.~A., et al. \ 1999, \aj, 118, 2014 
\item [490] Wilkes, B.~J., et al.\ 2005, \apj, 634, 183
\item [491] Willick, J.~A., et al.\ 1990, \apj, 355, 393
\item [492] Willis, A.~G., et al.\ 1976, \aaps, 25, 453 
\item [493] Willott, C.~J., et al.\ 2003, \mnras, 339, 173 
\item [494] Wilman, R.~J.\ 1999, \apss, 266, 55 
\item [495] Windhorst, R.~A., et al.\ 1984, \aaps, 58, 1
\item [496] Woo, J.-H., \& Urry, C.~M.\ 2002, \apj, 579, 530 
\item [497] Woudt, P.~A., et al.\ 2004, \aap, 415, 9
\item [498] Wright, A., \& Otrupcek, R.\ 1990, PKS Catalog (1990), 0 
\item [499] Wright, A.~E., et al.\ 1991, \mnras, 251, 330 
\item [500] Wright, A.~E., et al.\ 1994, \apjs, 91, 111 
\item [501] Wright, A.~E., et al.\ 1996, \apjs, 103, 145
\item [502] Xiang, L., et al.\ 2006, \aap, 454, 729
\item [503] Yamada, T., et al.\ 1993, \apjs, 89, 57
\item [504] Ye, T.~S., et al.\ 1995, \mnras, 275, 1218 
\item [505] Young, L.~M.\ 1999, \aj, 117, 1758 
\item [506] Yuan, Q., et al.\ 1998, PPM Observatory, 17, 1 
\item [507] Yuan, Q., et al.\ 2003, \apjs, 149, 53 
\item [508] Zaggia, S., et al.\ 1999, \aaps, 137, 75
\item [509] Zampieri, L., et al.\ 2005, \mnras, 364, 1419 
\item [510] Zezas, A., et al.\ 2002, \apjs, 142, 239 
\item [511] Zhang, X., et al.\ 1993, \aaps, 99, 545 
\item [512] Zhou, X., et al.\ 2003, \aap, 397, 361 
\item [513] Zickgraf, F.-J., et al.\ 2003, \aap, 406, 535 
\item [514] Zickgraf, F.-J., et al.\ 2005, \aap, 433, 151 
\item [515] Zoonematkermani, S., et al.\ 1990, \apjs, 74, 181 

\end{itemize}

\end{document}